\documentclass[aps,prb,twocolumn,superscriptaddress]{revtex4-2}
\usepackage{amsmath, amssymb,  graphicx}
\usepackage[colorlinks=true ,urlcolor=blue,urlbordercolor={0 1 1}]{hyperref}
\usepackage{color}
\usepackage{xcolor}
\usepackage{braket}
\usepackage[utf8]{inputenc}
\usepackage{bm}
\usepackage{verbatim}
\usepackage{graphicx}
\usepackage{amsmath}
\usepackage{color}
\usepackage{float}
\usepackage{bbm}
\usepackage{amssymb}
\usepackage{slashed}
\usepackage{wasysym,bm,bbm,dsfont}

\begin{document}
\title{Strong pairing from small Fermi surface  beyond weak coupling: \\Application to La$_3$Ni$_2$O$_7$}

\author{Hui Yang }
\thanks{These two authors contributed equally}
\author{Hanbit Oh}
\thanks{These two authors contributed equally}
\author{Ya-Hui Zhang}
\email{yzhan566@jhu.edu}
\affiliation{William H. Miller III Department of Physics and Astronomy, Johns Hopkins University, Baltimore, Maryland, 21218, USA}

\date{\today}

\begin{abstract}
The studies of high-temperature superconductors raise a fundamental question: Can a small Fermi surface phase, which violates the Luttinger theorem, exist and give rise to superconductivity? Our work provides a positive answer through a controlled theory based on a bilayer model with strong inter-layer spin-spin coupling ($J_\perp$) but no inter-layer hopping ($t_\perp$). Then small hole doping of the rung-singlet insulator with two electrons per rung naturally leads to small hole pockets with Fermi surface volume per flavor smaller than the free fermion result by $1/2$ of the Brillouin zone(BZ). We construct a new t-J model on a bilayer square lattice, so called ESD t-J model and employ a generalized slave boson theory, which captures this small Fermi surface phase at small hole doping $x$. This metallic state is an intrinsically strongly correlated Fermi liquid beyond weak coupling theory, violating the perturbative Luttinger theorem but consistent with the Oshikawa's non-perturbative proof. 
We further show that it transitions into an inter-layer paired $s'$-wave superconductor at lower temperature through Feshbach resonance with a virtual Cooper pair, with a surprising doping-induced crossover from Bardeen-Cooper-Schrieffer (BCS) to Bose-Einstein condensation (BEC) at higher hole doping levels. 
This leads to a superconducting dome centered around $x=0.5$, with the normal state changing from the conventional Fermi liquid in the $x>0.5$ to the unusual small Fermi surface state in the $x<0.5$ side. Our theoretical findings including phase diagrams are also confirmed by density matrix renormalization group (DMRG) simulation in quasi one dimension. Applying our theoretical framework, we provide a plausible scenario for the  recently found nickelate La$_3$Ni$_2$O$_7$ materials.
 \end{abstract}

 \maketitle

 \section{Introduction}
Elucidating the mechanism of high temperature superconductor has been one of the central challenges in strongly correlated physics. The predominant theoretical efforts so far have hinged on the celebrated one-orbital single-layer Hubbard model or the standard t-J model inspired by the cuprate materials\cite{lee2006doping}. Despite of extensive studies, there is no well-established theory of the pairing mechanism in the high Tc cuprates. On the overdoped side (with larger hole doping $x$), both weak coupling theory based on spin fluctuation \cite{2002cond.mat..1140C} and strong coupling approach \cite{anderson1987resonating} such as the slave boson theory\cite{lee2006doping}  reach at least qualitative consensus. 
The situation in the underdoped regime remains controversial, due to the ambiguous nature of the normal state. There is experimental evidence for small Fermi surface when the hole doping $x$ is smaller than a critical value \cite{doi:10.1146/annurev-conmatphys-031218-013210}. Nevertheless, it remains unclear whether some long-range orders are responsible for Fermi surface reconstruction. There have been interesting proposals of a symmetric metallic phase violating the Luttinger theorem\cite{senthil2003fractionalized,mei2012luttinger,zhang2020pseudogap}, but the existence of such an exotic metallic phase remains a topic of debate.

The aim of our paper is to demonstrate a controlled theory of a small Fermi surface phase without symmetry breaking. The primary difficulty of the standard t-J model stems from the presence of localized spin moments from the Mott insulator. While it seems natural that doped holes into the Mott insulator at $n=1$ filling might form small hole pockets, the localized spin moments prefer to form antiferromagnetic order, further complicated by the hole motion. 
Based on this crucial observation, we detour our focus to a bilayer system, where an unambiguous theory of small Fermi surface state and superconducting instability is possible.
We here consider a bilayer model with strong inter-layer antiferromagnetic spin-spin coupling $J_\perp$ but no inter-layer hopping. Within the Mott insulator at $n=1$ filling per site (two electrons per rung of the bilayer lattice), the spin moments simply form a rung-singlet phase due to the large $J_\perp$. In the rung-singlet phase, the spin moments are gapped and any magnetic ordering is suppressed in the large $J_\perp$ regime. Then doped holes can naturally form small Fermi surfaces on top of the trivially gapped rung-singlet insulator. However, the Fermi surface volume per flavor is now smaller by half of the Brillouin zone compared to the free fermion model. This phase  violates the conventional Luttinger theorem, calling for a new strong coupling theory.

\begin{figure}[ht]
    \centering    \includegraphics[width=\linewidth]{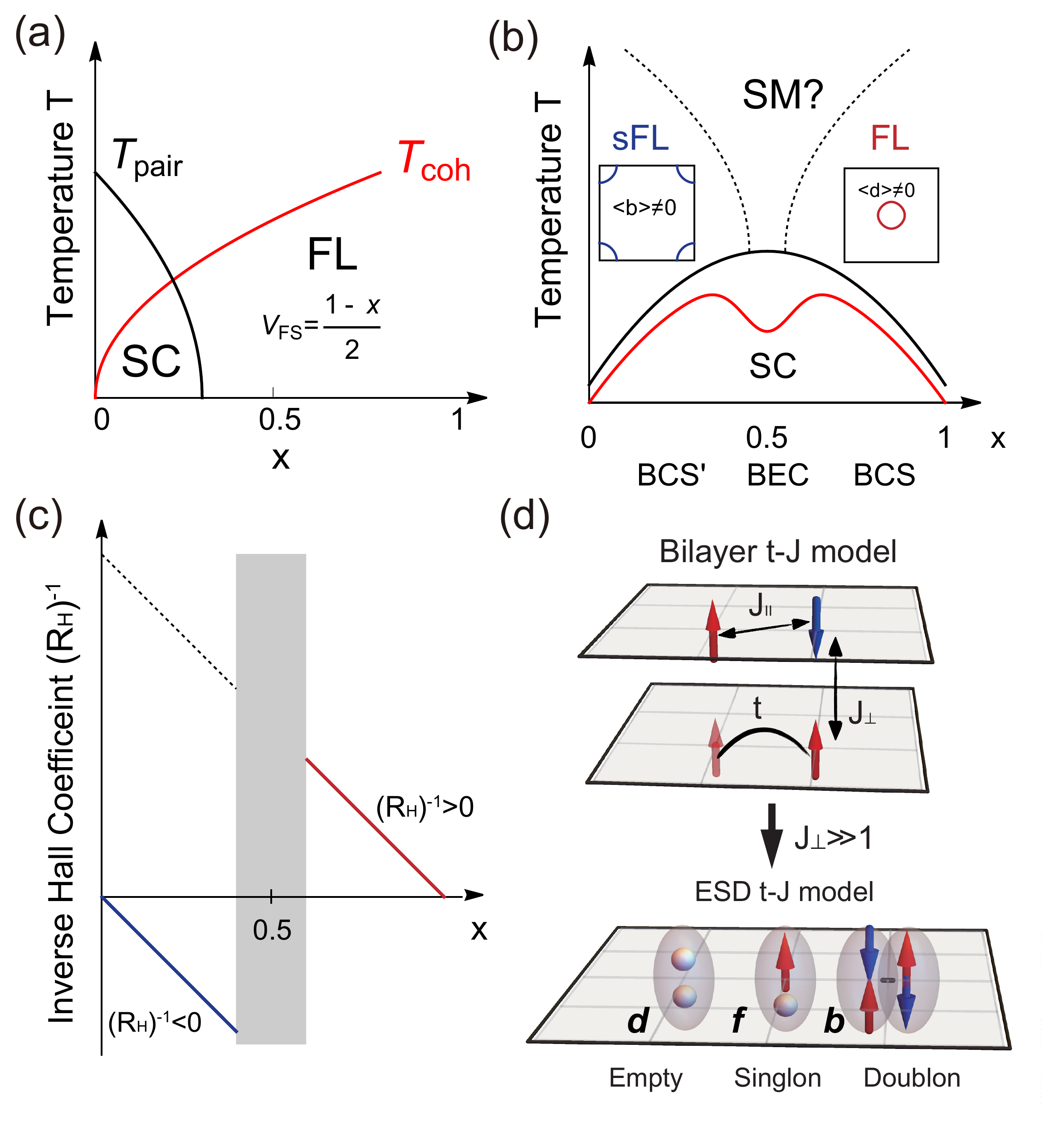}
    \caption{
    \textbf{(a-b) Illustrated phase diagram of standard one-orbital $t-J$ model from doping a single-layer spin 1/2 Mott insulator and the ESD t-J model.}
    The black and red lines indicate the pairing scale and the coherence scale.  
    FL (sF) stands for Fermi liquid (second Fermi liquid) respectively. SC (SM) is for  superconductor (strange metal).  
    (a) In the conventional model (i.e. cuprate), the pairing strength always decreases with the hole doping with the pairing disappearing before $x=0.5$.
    (b) In our ESD model, there is a superconducting pairing dome around $x=0.5$, which separates two different normal states above $T_c$. In the inset of the sFL and FL phase we show small hole pocket around $\vec k=(\pi,\pi)$ and small electron pocket around $\vec k=(0,0)$ respectively.
    At low temperature, both the sFL and FL phase have inter-layer $s'$-wave pairing instability mediated by a virtual Cooper pair. There is also doping induced BCS to BEC crossover as we move towards $x=0.5$.  We dub the $x \simeq 0$ side as in the BCS$^\prime$ limit given that its normal state is a sFL phase with small Fermi surface.
  \textbf{(c) The schematic picture of  
the inverse of Hall coefficeint  with the doping $x$}. The red line is expected in the free fermion model while the blue line is anomalous and indicates the strongly correlated sFL phase.  
\textbf{(d) The illustration of ESD t-J model.} Due to the large $J_{\perp}$, the effective Hilbert space is simplified as the six states namely one empty, four singlon and one doublon states combining two layer degree of freedom. In our model, we call these six states as $d$, $f$, and $b$ state, respectively.
}
\label{fig:global_phase_diagram}
\end{figure}

In this paper, we propose a new model, so called the \textit{ESD t-J model}, to capture the essential physics.
In the large $J_\perp$ regime, we can simplify our analysis by focusing on just six states per rung (combining two layers together) namely, one empty, four singlon, and one doublon states (see Fig.~\ref{fig:global_phase_diagram}(d)). 
Through a generalized slave boson theory, our model naturally identify two distinct normal states : (I) a conventional Fermi liquid (FL) phase with a Fermi surface volume $A_{FS}=\frac{1-x}{2}$ per flavor for $x>0.5$, (II) an unconventional Fermi liquid phase with a Fermi surface volume $A_{FS}=-\frac{x}{2}$ for $x<0.5$ (See Fig.~\ref{fig:global_phase_diagram}(b)).  Here $x$ is the hole doping and the total density per rung is $n_T=2-2x$, with the filling per spin-layer flavor as $\nu=\frac{1-x}{2}$.
Notably, the small hole pocket state near $x=0$, dubbed as second Fermi liquid (sFL), can be viewed as a symmetric pseudogap metal because its Fermi surface volume is shifted by $1/2$ per flavor compared to the FL phase due to partial Mott localization of one charge carrier per site (per layer), similar to  certain theories of the underdoped cuprate\cite{zhang2020pseudogap}. However, now it does not accompany any symmetry breaking or fractionalization, representing a generalization of symmetric mass generation (SMG)\cite{wang2022symmetric,you2018symmetric,PhysRevB.107.195133} away from integer filling.
The existence of the sFL phase challenges the traditional definition of a Fermi liquid through adiabatic evolution from the non-interaction limit. However, it is consistent with the non-perturbative proof of the Luttinger's theorem by Oshikawa\cite{oshikawa2000topological} in this model. The sFL phase is an intrinsically strongly correlated Fermi liquid existing only in the regime with strong four-fermion interaction. It is obviously beyond any conventional mean field theory with only bilinear term of electron operator, but can be easily captured by our slave boson theory. Despite the absence of  non-interacting limit, it is still a Fermi liquid phase in the sense that there is a finite quasi-particle residue $Z$ for excitations around the Fermi surface.

We also provide a theory of superconducting instability from the small Fermi surface sFL phase at lower temperature. Interestingly the low energy physics is described by an emerging fermion-boson model, similar to the Feshbach resonance discussed in cold atoms \cite{PhysRevLett.93.250402,PhysRevA.65.053607,PhysRevLett.105.195301}. Feshbach resonance-induced pairing has been explored in previous studies in other contexts \cite{doi:10.1126/sciadv.abh2233, PhysRevLett.131.056001}, though from totally distinct pairing mechanism. In our model, both the sFL and FL phases are unstable to inter-layer s$^\prime$-wave pairing. Starting from the under-doped region with $x=0$ or $x=1$, as doping level $x$ approaches the middle around $x\approx 0.5$, the energy of the virtual Cooper pair must decrease due to filling constraint, leading to a Feshbach resonance and a BCS to BEC crossover. Remarkably, the overdoped regime at $x\approx 0.5$ falls into the BEC regime, while the underdoped region with $x$ near 0 or 1 remains in the BCS regime. This provides a completely new phase diagram, as illustrated in Fig.~\ref{fig:global_phase_diagram}(b), in stark contrast to the cuprates' phase diagram

We propose that the recently discovered nickelate superconductor La$_3$Ni$_2$O$_7$ under high pressure \cite{sun2023signatures} is an ideal platform to explore the sFL phase and the doping induced Feshbach resonance. There have already been numerous experimental \cite{liu2023electronic,hou2023emergence,zhang2023high,yang2023orbital,zhang2023effects} and theoretical\cite{luo2023bilayer,zhang2023electronic,yang2023possible,sakakibara2023possible,gu2023effective,shen2023effective,wu2023charge,christiansson2023correlated,liu2023s,cao2023flat,qu2023bilayer,lu2023superconductivity,jiang2023pressure,tian2023correlation,zhang2023strong,qin2023high,huang2023impurity,zhang2023trends,jiang2023high,yang2023minimal,qu2023bilayer,2023arXiv230809044Q,2023arXiv230807386Z,2023arXiv230812750K,2023arXiv230811614J} studies of the bilayer nickelate.
One key observation is that the Hund's coupling can share the large $J_\perp$ of the $d_{z^2}$ orbital to the $d_{x^2-y^2}$ orbital \cite{oh2023type,lu2023interlayer}, thus the essential physics is described by a bilayer type II t-J model with strong $J_\perp$ but no $t_\perp$ if the $d_{z^2}$ orbital is Mott localized\cite{zhang2020type,zhang2021fractional,zhang2022pair}. We will show that a similar ESD t-J model is derived in the large $J_\perp$ limit and thus we expect the same physics in the nickelate system as described above. 
To go beyond the mean-field theory, we simulate the bilayer type II t-J model using density matrix renormalization group (DMRG) in quasi one dimension. The results confirm that there is a pairing dome around $x=0.5$, and support the existence of  the sFL and FL phases by adding a large $V$.
 The previous studies based on bilayer model with a $J_\perp$ term \cite{oh2023type,lu2023interlayer,lu2023superconductivity} predict that the pairing gap decreases with $x$ implying that the optimal hole doping should be much smaller than $x=0.5$. Also these previous studies in the conventional electron operator basis obtain a conventional Fermi liquid with large Fermi surface as the normal state in the entire range of $x$. In this work we show that this conclusion is wrong for $x<0.5$ side in the large $J_\perp$ regime due to the non-perturbative effect of the $J_\perp$ coupling not captured by the previous approaches. Instead, within the reduced ESD t-J model in the large $J_\perp$ regime, we find a sFL phase with small Fermi surface when $x<0.5$ and its pairing scale increases with $x$ with a dome structure centered around $x=0.5$. To our best knowledge, this is the first controlled theory of superconductivity from a strongly coupled symmetric small Fermi surface normal state. The current experiment in nickelate La$_3$Ni$_2$O$_7$ is limited only at $x=0.5$. Applying our theoretical findings to the experiments, we offer a plausible explanation of the puzzle why a high $T_c$ superconductor with $T_c=80$ K can be found at such a substantial doping level of $x=0.5$. Future experiments with varying $x$ can test our prediction of two different normal states in the two sides, indicated by a rapid jump of the Hall number (see Fig.~\ref{fig:global_phase_diagram}(c)).

\section{Low energy effective model: the ESD t-J model}
\label{secII}
In this section, we introduce the minimal theory which captures the essential physics. The effective model can be derived from the large $J_\perp$ limit of an one-orbital $t-J_\perp-V$ model on the bilayer square lattice, 
\begin{align}
    H=&-t \sum_{l,\sigma,\langle ij \rangle}Pc^\dagger_{i;l;\sigma}c_{j;l;\sigma}P+J_\parallel \vec{s}_{i;l}\cdot\vec{s}_{i;l}\nonumber\\
    &+J_\perp \sum_i \vec s_{i;t}\cdot \vec s_{i;b}+V\sum_{i} n_{i;t}n_{i;b}
    \label{eq:one_orbit_t_J}
\end{align}
where $P$ is the projection operator to forbid the double occupancy. 
$l = t, b$ labels the layer, and $i,j$ is for the site.  The model has a $(U(1)_t\times U(1)_b\times SU(2)_S)/Z_2$ symmetry with separate charge conservation in the top and bottom layers. We also assume a mirror reflection symmetry $\mathcal M$ which exchanges the two layers. Similar model has been studied by one of us in the context of moir\'e systems with valley playing the role of spin\cite{zhang2020spin}. The model has also been discussed in the context of bilayer optical lattice\cite{bohrdt2022strong,hirthe2023magnetically}.

In the large $J_\perp$ limit, we can use a simpler mode dubbed as the \textit{ESD t-J model}. 
Here \textit{ESD} means that we have an empty state, a singly occupied state, and doubly occupied state at each rung (combining two layers). In this notation, the standard t-J model is called ES t-J model as the doubly occupied state is forbidden there. Although we derive the ESD-t-J model from the specific one-orbital bilayer model in Eq.~\ref{eq:one_orbit_t_J}, we believe it is quite generic 
 for the physics in the large $J_\perp$ regime and does not care about details.

\begin{figure}
    \centering
\includegraphics[width=0.44\textwidth]{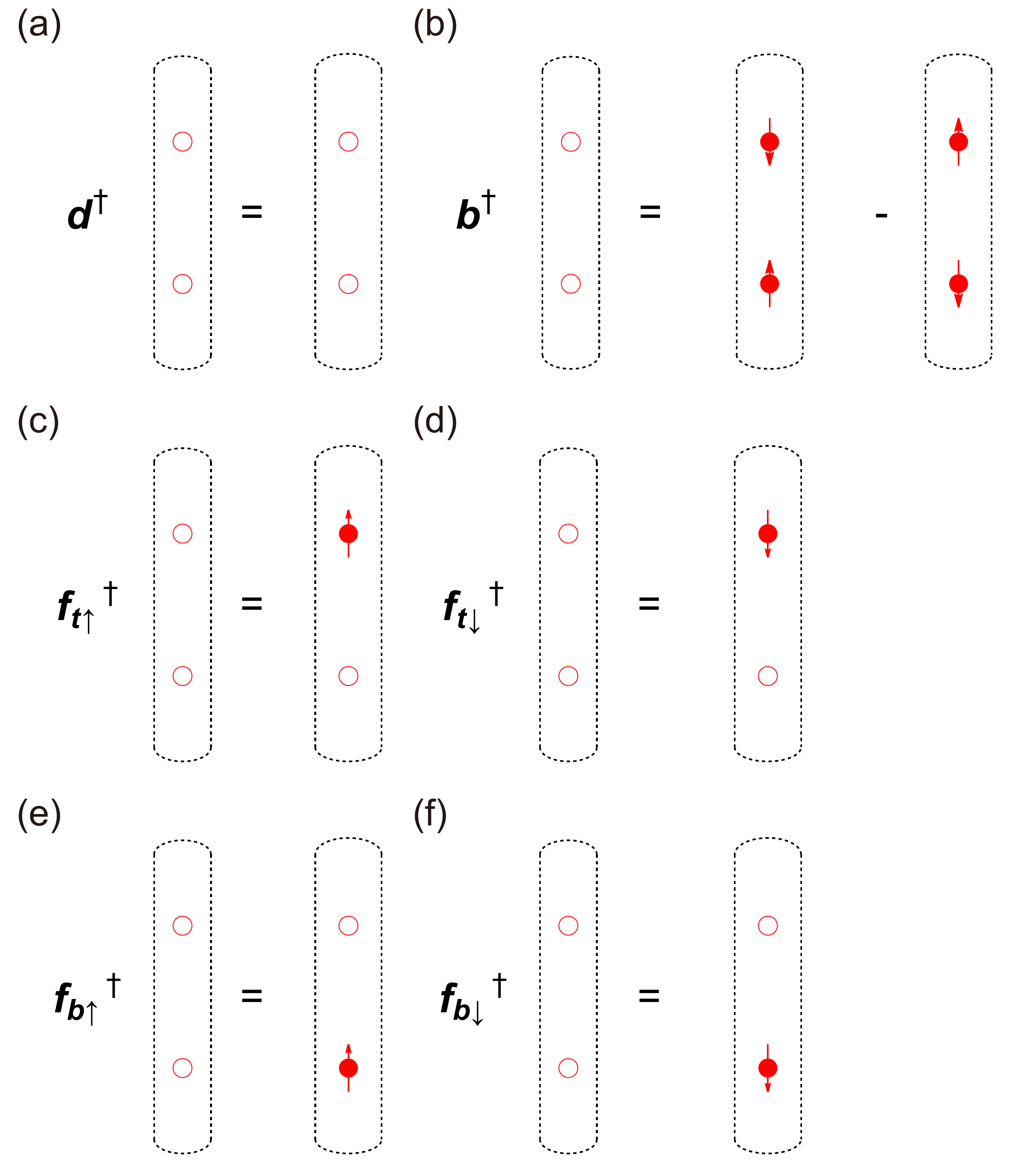}
    \caption{
    \textbf{Illustrated six states in the ESD-t-J model.} 
    Each state is defined by combining the top and bottom layer, as boxed together. 
    The red empty (solid) circles represent the holes (filled electrons) within the one-orbital.  
    The arrow is depicting the spin-1/2 moment of the electron. 
    (a-b) The $b^\dagger$ and $d^\dagger$ create rung-singlet doublon state and an empty state, respectively, while (c-f) $f^\dagger_{l\sigma}$ create fermion with carrying 1/2 spin $\sigma=\uparrow,\downarrow$ at either top, bottom layer $l=t,b$.}
    \label{fig:partons}
\end{figure}

In the large $J_\perp$ limit, we can solve the two-site problem at each $i$ first and reduces the local Hilbert space with only  $6$ states per site, combining two layers together (see Fig.\ref{fig:partons}).
The six states can be labeled as as follows.  
First, the doublon state $\ket{b}=[ c_{b;\downarrow}^\dag
c_{t;\uparrow}^\dag
  -
  c_{b;\uparrow}^\dag
c_{t;\downarrow}^\dag
]|G\rangle /\sqrt{2}$, which is just a rung singlet. $|G\rangle$ denotes an vacuum state. 
The second state  is just an empty state $\ket{d}=\ket{G}$. For the singlon state with one hole per rung, we have four states carrying a total spin $S=1/2$, labeled as $\ket{l,\sigma}=c^{\dag}_{l;\sigma} |G\rangle$ 
with $l=t,b$ and $\sigma=\uparrow,\downarrow$. 
Within this restricted Hilbert space, the electron operator can be projected in a compact form,  
\begin{eqnarray}
c_{i;l,\sigma}&=& |d\rangle_{i} \langle l,\sigma |_{i}
+\frac{\epsilon_{\sigma \sigma'}}{\sqrt{2}}|
\overline{l},\sigma' \rangle_{i}
\langle b|_i ,
\end{eqnarray}
with introducing the antisymmetric tensor 
and $\epsilon_{\uparrow \downarrow}=-\epsilon_{\downarrow \uparrow}=1$, 
the opposite layer index where $\overline{l}$, i.e. $\overline{t}=b, \overline{b}=t$.
Subsequently, the new model formulated in this $1\oplus 1\oplus 4$=6 dimensional Hilbert space becomes 
\begin{align}
    H=-t\sum_{l;\langle ij\rangle}
    P c_{i;l;\sigma}^\dag c_{j;l;\sigma} P+
\sum_{i} \epsilon(n_{d;i} +n_{b;i}),
\label{esd_one_orbital}
\end{align}
where density operators, $n_{i;d}=\ket{d}_i\bra{d}_i$, $n_{i;b}=\ket{b}_i\bra{b}_i$, $n_{i;f}=\sum_{l;\sigma}\ket{l;\sigma}_i\bra{l;\sigma}_i$ are introduced. $P$ is the projection operator to the 6-states subspace. At each site we have $n_{i;d}+n_{i;f}+n_{i;b}=1$. On average we have $ \frac{1}{2}n_f+n_d=x$. 
Here, the averaged onsite energy of $d$ and $b$ is $\epsilon=(V-\frac{1}{4}J_\perp)/2$. 
$-\epsilon$ can be viewed as the binding energy of the Cooper pair. If $\epsilon$ is negative and large, then the single electron is gapped and the doped holes (electrons) form tightly bound pairs close to the $x=0$ ($x=1$) region.  The physics of large and negative $\epsilon$ can be captured by a simple hard-core boson model (or equivalently XY model) with only $\ket{d}$ and $\ket{b}$ in the Hilbert space. In the $\epsilon \rightarrow -\infty$ limit, the electron operator is gapped and only the Cooper pair operator $\Delta_i^\dagger=\ket{b}_i\bra{d}_i$ matters. This is in the strong BEC limit with the $T_c$ determined by the phase stiffness. We expect $T_c \sim \frac{t^2}{|\epsilon|}$, which is the effective hopping of the on-site Cooper pair..

In this paper we are interested in the realistic regime with  negative but small $\epsilon$ or positive $\epsilon$, so single electron or hole should be kept in the low energy theory. Superconductivity, if possible, should be understood in the BCS picture and the Fermi surface of the normal state is important. We will develop an analytical framework to deal with this regime in the next section.

\section{Phase diagram from a generalized slave boson mean field}
\label{secIII}
In the ESD t-J model, we remove the strong four-fermion interaction $J_\perp$ with the cost of a restricted Hilbert space, which can then be conveniently dealt with using a generalized slave boson representation,
\begin{eqnarray}
c_{i;l,\uparrow}&=& d^\dag_{i} f_{i;l,\uparrow}+\frac{1}{\sqrt{2}}
f_{i,\overline{l},\downarrow}^\dag b_i ,\\ 
c_{i;l\downarrow}&=& d_i^\dag f_{i;l,\downarrow}-\frac{1}{\sqrt{2}}
f_{i;\overline{l},\uparrow}^\dag b_{i},
\label{parton_one}
\end{eqnarray}
where $\overline{l}$ is an opposite layer index. We have a local constraint $n_{i;b}+n_{i;d}+n_{i;f}=1$, and $n_f+2n_d=2x$ on the average.  $b,d$ are slave boson operators which annihilate empty and doublon states, while $f_{i;l \sigma}$ is the usual Abrikosov fermion which annihilates the spin-1/2 singlon states. As in the usual slave boson theory\cite{lee2006doping}, there is a U(1) gauge redundancy:
$f_{i;l,\sigma}\rightarrow e^{i\theta_i} f_{i;l,\sigma}$, $d_{i;l,\sigma}\rightarrow e^{i\theta_i} d_{i;l,\sigma}$, and $b_{i;l,\sigma}\rightarrow e^{i\theta_i} b_{i;l,\sigma}$, which introduces an emergent U(1) gauge field $a_\mu$.  Meanwhile, the global $U(1)_c$ symmetry transformation $c_{i;l \theta} \rightarrow e^{i \theta_c} c_{i;l \theta}$. Note this $U(1)_c$ symmetry is a combination of the $U(1)_t$ and the $U(1)_b$ symmetry and corresponds to the total charge conservation. We will ignore the $U(1)_l$ symmetry corresponding to the relative charge between the two layers. We assign the $U(1)_c$ global symmetry in the following way: $f_{i;l\sigma}\rightarrow f_{i;l \sigma}, d_i \rightarrow e^{-i\theta_c} d_i, b_i \rightarrow e^{i \theta_c} b_i$.  We also introduce a probing field $A_\mu$ for this $U(1)_c$ global symmetry.  In the end, we have $f_{l \sigma}$ couples to $a_\mu$, $d$ couples to $-A_\mu+a_\mu$ and $b$ couples to $A_\mu+a_\mu$. In the ansatz we will discuss in this paper, the emergent gauge field $a_\mu$ is always higgsed.  So we only target featureless phases without any fractionalization. However, some of these seemingly conventional phases are beyond conventional mean-field frameworks without using the slave boson theory.

Substituting the parton to the ESD-t-J model, we rewrite it in the form,
\begin{eqnarray}
   H \!\!&=&\!\!-t\sum_{l,\langle i,j\rangle}\!\!
\left[f^\dag_{i,l,\uparrow}d_{i} +\frac{b_i^\dag f_{i,\overline{l},\downarrow}}{\sqrt{2}}
\right]\!\!
\left[d^\dag_{j} f_{j;l,\uparrow}+\frac{f_{j,\overline{l},\downarrow}^\dag b_{j}}{\sqrt{2}}
\right]\notag\\
&&+
\left[f^\dag_{i,l,\downarrow}d_{i} -\frac{b_i^\dag f_{i,\overline{l},\uparrow}}{\sqrt{2}}
\right]
\left[d^\dag_{j} f_{j;l,\downarrow}-\frac{f_{j,\overline{l},\uparrow}^\dag b_{j}}{\sqrt{2}}
\right]\notag\\
&&+\sum_{i}
\delta_{f} n_{f;i}
+\delta_{d} n_{d;i}
+\delta_{b} n_{b;i}.
\end{eqnarray}
with $\delta_{f}=-(\mu_{0}+\mu)$, 
$\delta_{d}=\epsilon-(\mu_{0}+2\mu)$,  $\delta_{b}=\epsilon-\mu_{0}$, and $\epsilon=(V-\frac{1}{4}J_\perp)/2$. 
Two chemical potentials are introduced to as Lagrangian multipliers to conserve the local constraints, $n_{i;d}+n_{i;b}+n_{i;f}=1$,and $n_f+2n_d=2x$.

Introducing the four order parameters, $\chi_{l;i,j} \equiv \sum_{\sigma} \langle f_{i;l;\sigma}^{\dagger}f_{j;l;\sigma} \rangle$, $\Delta_{i,j}\equiv 
\langle f_{i;t;\uparrow}f_{j;b;\downarrow} -f_{i;t;\downarrow}f_{j;b;\uparrow} 
     \rangle$,$\langle d \rangle$, and $\langle b \rangle$, the mean field Hamiltonian can be expressed as,
\begin{eqnarray}
    H_{MF} &=& H_{f}+H_{d}+H_{b}\\
  H_{f}&=&-C_{f} \sum_{l,\langle i,j \rangle} f^{\dagger}_{i;l;\sigma}f_{j;l;\sigma}+h.c.+\sum_{i}\delta_{f}n_{f;i}\notag\\
&& + D_{f} \sum_{\langle i,j \rangle }
\left[
f^{\dagger}_{i;t;\uparrow}f^{\dagger}_{j;b;\downarrow}
-f^{\dagger}_{i;t;\downarrow}f^{\dagger}_{j;b;\uparrow}
\right] +h.c. \notag\\
H_{d}&=& 
\sum_{i}
\lambda_{d} d_{i} 
+H.C. +\sum_{i}\delta_{d}n_{d;i}\notag\\
H_{b}&=& 
\sum_{i}\lambda_{b} b_{i}+ h.c.
+\sum_{i}\delta_{b}n_{b;i} \notag
\end{eqnarray}
with parameters $C_{f},D_{f},\lambda_d,\lambda_b$'s (see Appendix \ref{secA1} for details).   After solving the mean-field Hamiltonian self-consistently, we obtain the hole doping dependence of the order parameters.  The pairing of the $f$ fermion is always in the inter-layer paired $s'$-wave channel with a $\cos k_x+\cos k_y$ dependence in momentum space.

\begin{figure}[tb]
    \centering
    \includegraphics[scale=0.46]{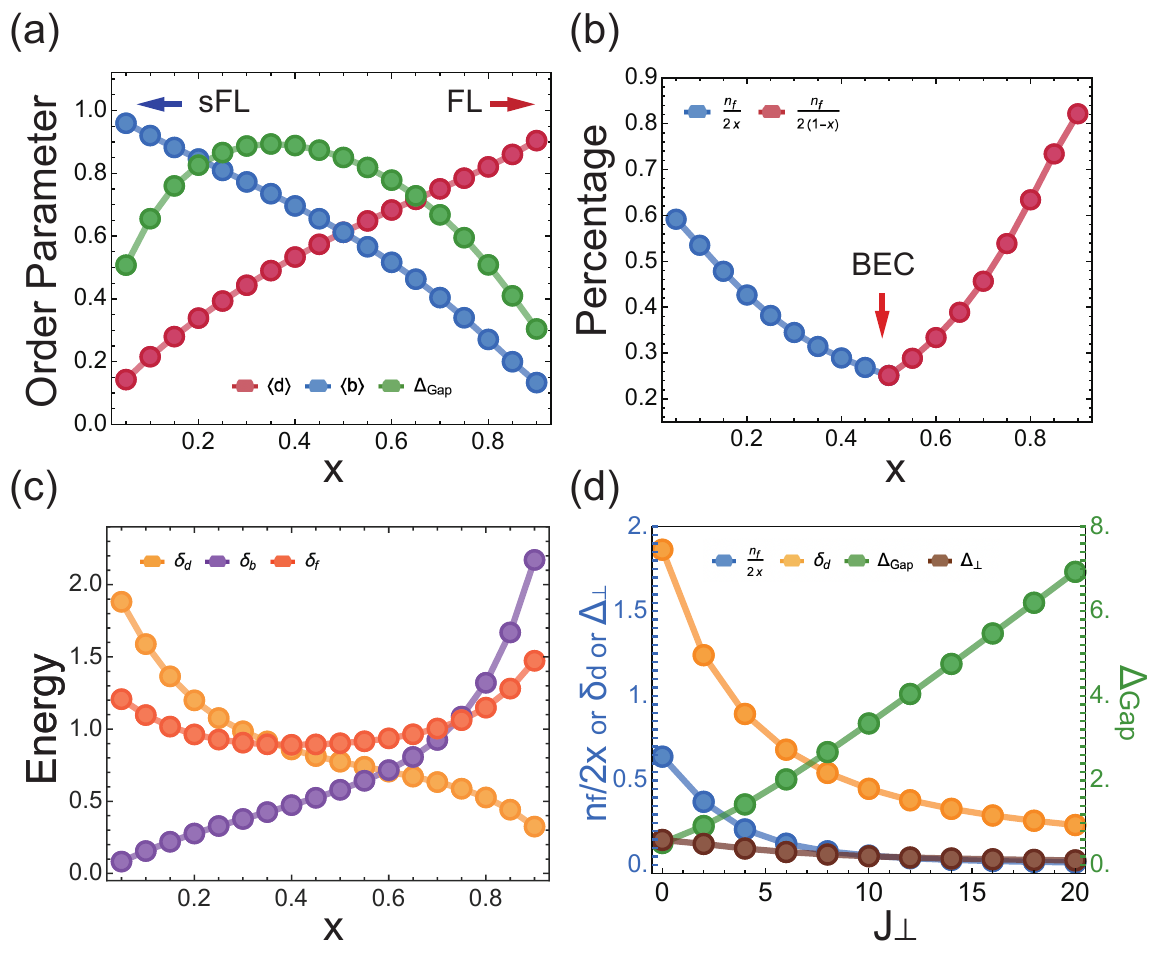 }
    \caption{\textbf{Slave boson mean-field results of ESD t-J model.} (a) Doping dependence of order parameters $\langle d\rangle, \langle b \rangle$ and energy gap $\Delta_{\mathrm{Gap}}$ at $J_{\perp}=1$, $V=0.3$. Here, the pairing gap $\Delta_{\mathrm{Gap}}$ is defined as a minimum gap of the Bogoliubov quasiparticle in $H_f$ in the momentum space.
 Above a small temperature scale,  $\langle b \rangle \neq 0, \langle d \rangle=0$ at $x$ close to 0 and $\langle d \rangle \neq 0, \langle b \rangle =0$ close to $x=1$, leading to the sFL phase and the FL phase in the two sides respectively. In lower energy scale, we have $d$ ($b$) further condensed in the two sides, resulting in the pairing instability of the sFL and the FL phase.
The optimal pairing gap appears near $x\simeq 0.5$. 
(b) The percentage of single hole state $\frac{n_{f}}{2x} (\frac{n_{f}}{2(1-x)})$ at $x<0.5$ ($x>0.5$). 
The single hole percentage is minimized near $x=0.5$, suggesting that the density of a single hole (single electron) is depleted by Cooper pair, a signature of the BEC limit with tightly bound Copper pair as the main carrier.
(c) The value of $\delta_{f},\delta_{d},\delta_{b}$ for showing the energy cost of the virtual cooper pair. In the small $x$ side, $b$ is condensed at high temperature scale. $d$ is the virtual cooper pair and its energy $\delta_d$ decreases when increasing $x$ towards $x=0.5$. Similarly, in the $x$ close to $1$ side, $d$ is condensed in the high temperature. Now $b$ is the virtual cooper pair and its energy cost $\lambda_b$ moves down when decreasing $x$ towards $x=0.5$.
(d) The single-hole percentage ($\frac{n_f}{2x}$), gap of the virtual Cooper pair ($\delta_{d}$), the order parameter ($\Delta_{\perp}\equiv 
\langle f_{i;t;\uparrow}f_{j;b;\downarrow} -f_{i;t;\downarrow}f_{j;b;\uparrow} 
     \rangle$) and the single particle gap $\Delta_{\mathrm{Gap}}$ at $x=0.1$. The dependence of $J_{\perp}$ shows that when $J_{\perp}$ is larger, the system is moving towards the BEC limit because the virtual Cooper pair energy $\delta_d$ decreases and the singly occupied state $n_f$ is depleted. The single-particle gap $\Delta_{\mathrm{Gap}}$ increases with $J_\perp$, although the mean field pairing term $\Delta_\perp$ decreases, which indicates that the single-electron gap is decided by the chemical potential instead of the pairing term in the BEC limit.}
\label{fig:mft_part1}
\end{figure}

Our main findings shown in Fig.~\ref{fig:mft_part1} are summarized as follows.
First, we check that two distinct Fermi-Liquids in the two opposite limits are dominated by $\langle b \rangle\neq 0$, and $\langle d \rangle\neq 0$ at $x=0$ and $x=1$ limit respectively. The two Fermi liquids have a BCS pairing instability at lower temperature scale with $d$ (or $b$) further condensed in the $x<0.5$ (or $x>0.5$) side, leading to a superconductor phase with inter-layer paired $s'$-wave pairing symmetry ($\cos k_x+\cos k_y$ in momentum space).
In the intermediate doping, both $\langle b \rangle$ and $\langle d \rangle$ condensate together at roughly the same energy scale, while the  the density of the singlon state is highly suppressed near $x=0.5$.
This  indicates the BEC limit where the Copper pair is tightly bound and these paired electrons (holes) become the main carriers.  
To elucidate the physics, we further analyze the energy cost of the paired hole (electron), $\delta_{d}$ ($\delta_b$) which goes down when moving $x$ towards the $x=0.5$ region. It implies that the virtual cooper pairing is energetically favored and causes the Feshbach resonance around $x=0.5$, so there is a doping controlled BCS to BEC crossover from the two sides to the middle.
Finally, we find larger $J_\perp$ moves the system towards the BEC limit while larger $V$ moves the system towards the BCS limit, as illustrated in Fig.~\ref{fig:mft_part1} (d). 
In the following, we discuss the main physics in the $x<0.5$ and in the $x>0.5$ side in more details respectively.

\subsection{sFL phase and Oshikawa-Luttinger theorem}
\label{secIIIA}

\begin{figure}[tb]
    \centering
\includegraphics[scale=0.6]{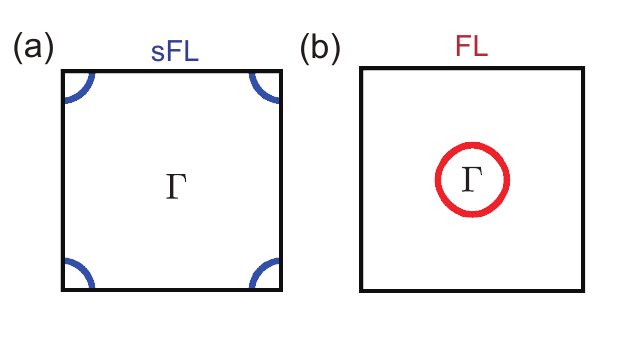}
    \caption{\textbf{Illustration of the Fermi surfaces of (a) sFL phase with $b$ condensed and $d$ gapped; and (b) FL phase with $d$ condensed and $b$ gapped.}
Two specific cases, $x=0.15$, $x=0.85$ with $J_{\perp}=1$, $V=0.3$ are chosen, based on our slave-boson mean-field calculation. 
(a) While the hole pockets are positioned at $M$ point for sFL, (b) electron pockets are centered $\Gamma$ point for FL.   Either the sFL or FL phase has four pockets due to spin-layer four fold degeneracy. The sFL and FL phases are stable only at finite temperature above the BCS critical temperature.
}
    \label{fig:fermi-surface}
\end{figure}

As illustrated in Fig.~\ref{fig:fermi-surface}, there are two distinct Fermi liquid phase in the ESD t-J model. In our slave boson theory, they correspond to the condensation of $b$ and $d$ bosons respectively.  At small $x<0.5$, at finite temperature we have $\langle b \rangle \neq 0$, while $d$ is gapped. Note that at $x=0$ the ground state is just a product state of the $\ket{b}$ state. Then at finite but small $x$, the doped holes just form small hole pockets with Fermi surface volume $A_{FS}=-\frac{x}{2}$ per flavor on top of the background $\ket{b}$ state. This Fermi surface area is smaller than free fermion result ($A_{FS}=\frac{1-x}{2}$) by half of the BZ. We dub this unconventional metallic phase as second Fermi liquid (sFL). Its existence is quite obvious in our ESD t-J model when the energy offset $\epsilon$ is positive and large. In this case the $\ket{d}$ state is gapped at small $x$. We are then left with a simple four-flavor $t-J$ model formed by the $\ket{b}$ and four singlon states, but with $n_{b}=1-2x$ and $n_f=2x$, so it is in the extremely overdoped side compared to the conventional slave boson theory\cite{lee2006doping}. Then naturally we just get a Fermi liquid with $\langle b \rangle \neq 0$ and $f$ forms four small hole pockets.

This sFL phase is beyond any weak coupling theory of the original model in Eq.~\ref{eq:one_orbit_t_J} as it is disconnected to the non-interacting limit ($J_\perp=0$). But it is allowed by the Oshikawa's non-perturbative proof of the Luttinger theorem. The conclusion follows from the symmetry $\big(U(1)_t \times U(1)_b \times SU(2))/Z_2$ and a mirror reflection symmetry $\mathcal M$ which exchanges the two layers in the bilayer model as in Eq.~\ref{eq:one_orbit_t_J}. The only role of the mirror reflection symmetry $\mathcal M$ is to guarantee that the two layers have the same density and do not play any other important role. We derive the conclusion following the Oshikawa flux threading \cite{oshikawa2000topological}, which offers a non-perturbative constraint that does not rely on any detailed description of the phase.

We restrict our focus to symmetric and featureless Fermi liquids, characterized by the absence of fractionalization, where the phase is well described by the Landau Fermi liquid theory. The total number of electrons per site (summed over layer and spin) is $n_T=2(1-x)$ and the density per flavor is then $\nu=\frac{1-x}{2}$. In the usual case, one can thread flux corresponding to a U(1) symmetry of each flavor, restricting the Fermi surface volume $A_{FS}=\nu$ mod 1 in units of the total area of the Brillouin zone (BZ) \cite{oshikawa2000topological}. Notably we now only have three U(1) global symmetries, with four flavors.  We can thread a flux corresponding to $U(1)_t$, this will lead to $A_{FS; t\uparrow}+A_{FS;t\downarrow}= 2\nu$ mod 1.  Similarly, threading a flux of $U(1)_b$ leads to $A_{FS; b\uparrow}+A_{FS;b\downarrow}= 2\nu$ mod 1. Combining rotation symmetry and the mirror refection leads $A_{FS;b\uparrow}=A_{FS;t\downarrow}=A_{FS;b\uparrow}=A_{FS;b\downarrow}=A_{FS}$. Consequently, two solutions satisfying the above constraints are: (I) $A_{FS}=\nu$ mod 1 and (II) $A_{FS}=\nu-\frac{1}{2}$ mod 1.

The lack of U(1) symmetry for each flavor surprisingly leads us to conclude that there exist two distinct classes of Fermi liquids. This can be seen as a generalization of the $Z_2$ classification of band insulators \cite{PhysRevLett.95.226801} to the symmetry-enriched Fermi liquid. The first class with $A_{FS}=\nu=\frac{1-x}{2}$ is consistent with the conventional Fermi liquid connecting to the free fermion description with $J_\perp=0$.  On the other hand, the second class with $A_{FS}=\nu-\frac{1}{2}=-\frac{x}{2}$ in the small $x$ limit has a small hole pocket, resembling a symmetric pseudogap metal. We emphasize that the state does not accompany any symmetry breaking or fractionalization, even though it possesses a small hole pocket smaller by half than the Fermi liquid as proposed in the underdoped cuprate. We name this novel phase as the second Fermi liquid (sFL). The sFL phase cannot be described by any mean field theory with  bilinear Hamiltonian of electron operator, and requires the inclusion of a strong four-fermion interaction $J_\perp$ term in Eq.~\ref{eq:one_orbit_t_J}. Therefore, it cannot be adiabatically connected to the conventional Fermi liquid (FL) phase, a phase transition between them is bound to happen. 

We note that it is very challenging (if not impossible) to capture the sFL phase directly within the bilayer one-orbital t-J model in Eq.~\ref{eq:one_orbit_t_J}. For example, we can use a conventional slave boson theory $c_{i;a;\sigma}=b^\dagger_{i;a}f_{i;a\sigma}$ to study the phase diagram of the bilayer one-orbital t-J model\cite{oh2023type,lu2023interlayer}. Then one can indeed obtain an inter-layer s-wave paired superconductor, but its normal state is the FL phase with large Fermi surface. The sFL phase with small Fermi surface is missed in this approach because the non-perturbative effect of $J_\perp$ is beyond the simple mean field theory directly in Eq.~\ref{eq:one_orbit_t_J}. In contrast, here we handle the large $J_\perp$ by reducing the model to the ESD t-J model and then a simple slave boson mean field theory describes the sFL phase quite easily. As we will see, in our theory the sFL phase also has an instability of inter-layer s-wave pairing, but with a momentum dependence $\cos k_x+\cos k_y$.

\subsection{BCS$^\prime$-BEC crossover from the sFL phase for $x<0.5$}
\label{secIIIB}
In the limit of small $x$, the system can be viewed as doped away from rung-singlet Mott insulator with $b$ as the vacuum. If we assume $d$ is gapped presumably with a large $V$ or at finite temperature, then only $b$ state is condensed $\langle b \rangle \neq 0$ with $n_{b}=1-2x$.
Then our original parton ansatz reduces to $c_{i;l\sigma}\sim \frac{\epsilon_{\sigma,\overline{\sigma}} }
{\sqrt{2}}
b f^\dagger_{\bar l, \bar \sigma}$.
Now we have $n_f=2x$ which attributes to small hole pockets with $A_{FS}=-\frac{x}{2}$ per flavor. Because $b$ couples to $A_\mu+a_\mu$, the condensation of $b$ locks $a_\mu=-A_\mu$. Then $f$ couples to $-A_\mu$ and $d$ couples to $-2A_\mu$. Consequently, one identify $f$ as a hole operator, and $d$ as a Cooper pair of holes. The nature of $f$ is consistent with the Fermi surface volume $A_{FS}=-\frac{x}{2}$ and the hole pocket is centered at $\vec{k}=(\pi,\pi)$

To further understand the pairing instability of the sFL phase, we construct an effective fermion-boson model by replacing the $b$ operator with its expectation value $\langle b \rangle$:
\begin{align}
H=\sum_{\langle i, j\rangle}-&g(d_i^\dag+d_j^\dag)(f_{i,b,\downarrow}f_{j,t,\uparrow}-f_{i,b,\uparrow}f_{j,t,\downarrow})+h.c.\notag\\
+&t_{\mathrm{eff}} f_{i,l,\sigma}^\dag f_{j,l,\sigma}+h.c.+\delta_d n_{d,i}\notag\\
-&tf_{i,l,\sigma}^\dag f_{j,l,\sigma}d_id_j^\dag+h.c.,
\label{eq:boson-fermion}
\end{align}
where $t_{\mathrm{eff}}=\frac{t\langle b\rangle^2}{2}$, $g=\frac{1}{\sqrt{2}}t\langle b\rangle$.
Interestingly, we find an emergent fermion-boson model similar to the physics of Feshbach resonance\cite{GURARIE20072,sheehy2007bec,PhysRevLett.96.060401}.  
The virtual cooper pair $d$ now mediates attractive interaction $V \sim - \frac{t^2}{\delta_d}$ and causes a BCS instability of the sFL phase. The critical temperature of the superconducting state inherited from the sFL phase is explicitly obtained as  (see the Appendix \ref{secA2}),
\begin{align}
T_{c}&
=\frac{16}{\pi}e^{\gamma}t_{\mathrm{eff}}\exp\left[ -\frac{\delta_{d}}{16 g^2 D(0)C_{\mathrm{FS}}}\right],
\label{eq:tc}
\end{align}
where the Euler constant $\gamma=0.577$ is used. 
We have a zero energy density of state $D(0)$, and 
the average of the form factor over the Fermi surface, $C_{\mathrm{FS}}=\langle (\cos k_x +\cos k_y)^2\rangle_{\mathrm{FS}}$. These quantities depends on the microscopics of the system, and we provide their doping dependence of Fig.\ref{fig:dos_dep} in Appendix \ref{secA2}.

Interestingly, we find that the energy offset $\delta_d$ decreases as increasing $x$ towards $0.5$ as shown in Fig.\ref{fig:mft_part1}(c). 
This is straightforward to understand. When $x>0.5$, we must have a finite number of $d$ states no matter how large its energy cost is because $n_f+2n_d=2x$ can not be satisfied anymore with $n_d=0$ when $x>0.5$.   Therefore, moving towards the $x=0.5$ regime, the virtual cooper pair $d$ state comes down and resonant with the Fermi surface, leading to a BEC to BCS crossover with the BEC limit in the overdoped regime with $x\approx 0.5$. We need to emphasize the BCS here should be dubbed as BCS$^\prime$ to highlight the unusual normal state with small Fermi surface.

\subsection{FL phase and BEC-BCS crossover for $x>0.5$}\label{secIIIC}

In the limit of $x$ close to $1$, we can start from the insulator of the $\ket{d}$ product state at $x=1$. 
Similar to the previous subsection, one can easily indicate that $d$ is condensed  $\langle d \rangle \neq 0$ with $n_d=2x-1$
Moreover, $f$ now becomes an electron operator, $c_{i;l,\sigma} \sim d^\dagger_i f_{i;l \sigma}$, while $b_i$ is now the virtual Cooper pair. 
Now we have electron density $n_f=2(1-x)$ which forms electron pocket with Fermi surface volume $A_{FS}=\frac{1-x}{2}$ centering at $\vec{k}=\Gamma$.
The similar boson-fermionic model as in Eq.(\ref{eq:boson-fermion}) can be achieved but coupled with dynamic $b$ field. The critical temperature $T_{c}$ for $x>0.5$ follows the similar form as given in Eq.\ref{eq:tc}, but now $\delta_{b}$ plays the role of the energy of the virtual Cooper pair.  Now we decrease $x$ towards the overdoped region at $x\approx 0.5$, the energy offset of $b$ must come down towards $x=0.5$ because we know $b$ must condense at $x<0.5$ side. 
Hence, we again find a BEC to BCS crossover but from the FL phase.

\section{Application to nickelate: type II t-J model}
\label{secIV}

 \begin{figure}[hb]
     \centering
\includegraphics[scale=0.42]{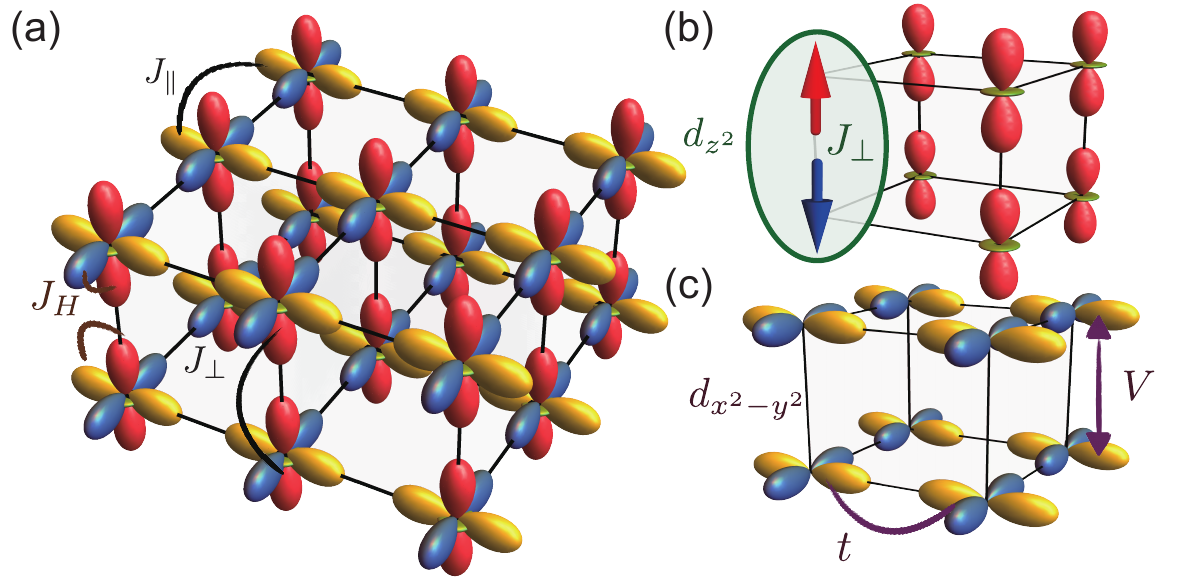}
     \vspace{-5pt}
     \caption{\textbf{Illustrated bilayer Kondo model in Eq.(\ref{eq:1}) of bilayer Nickelates.} 
    (a) Two distinct orbitals are living on the bilayer square lattice. 
    The $t,J$’s represent the hoppings and super-exchange interactions on square lattices. 
    $J_{H}$ is Hund orbital between two orbitals
     (b-c) We assume that the $d_{z^2}$ orbitals are localized and only provide spin-half moments forming a rung singlet, while the $d_{x^2-y^2}$ orbitals are mobile electrons, but with negligible inter-layer hopping and spin-spin coupling. The large $J_H$ will share the $J_\perp$ from the $d_{z^2}$ orbital to the mobile electron in the $d_{x^2-y^2}$ orbital automatically. There is no $J_\perp$ for the  $d_{x^2-y^2}$ orbital in our model.      }
     \label{fig:kondo_model}
 \end{figure}

In this section, we elaborate how our theory can be applied to the recently discovered bilayer nickelate La$_3$Ni$_2$O$_7$ under high pressure \cite{sun2023signatures}. 
We briefly review how to derive the type II t-J model for the doped spin-one Mott insulator with valence configuration $d^{8-x}$, as has been done in the previous paper \cite{oh2023type}.
Note that the density (summed over spin) per site is $n_{1}\simeq 1-x$ for $d_{x^2-y^2}$ orbital and $n_{2}\simeq 1$ for $d_{z^2}$ orbital, respectively. 
At $x=0$, we have one electron in the $d_{x^2-y^2}$ orbital and one electron in the $d_{z^2}$ orbital. They form a spin-one moment together due to a large Hund's coupling $J_H$. 
At finite $x$, holes prefer to be doped to one of the two orbitals in the general case with energy splitting between the two orbitals. 
The relative energy splitting of $d_{x^2-y^2}$ from $d_{z^2}$ one is $0.372$eV \cite{2023arXiv230606039S,2023arXiv230909462S}, and $d_{z^2}$ orbital is always filled providing a spin $1/2$ moment at each site for the entire range of $x$. 
It is useful to understand this situation by Kondo model. 
For example, at $x=0$ limit, a spin-half Mott insulator is formed by $d_{z^2}$ only, and $d_{x^2-y^2}$ orbital is empty. 
At generic $x$, we have itinerant electrons from the $d_{x^2-y^2}$ orbital coupling to the localized spin moments of $d_{z^2}$ orbital through the Hund's coupling $J_H$. 
The full Hamiltonian is then a bilayer Kondo model \cite{zhang2022pair} (see Fig.~\ref{fig:kondo_model}),
\begin{align}
    H&=- t \sum_{\langle ij \rangle} \sum_{l,\sigma} P c_{i;l;\sigma}^\dagger c_{j;l;\sigma}P +J_{\parallel} \sum_{\langle ij \rangle} \vec S_{i;l;c}\cdot \vec S_{j;l;c} \notag \\ 
    &+J_\perp \sum_i \vec S_{i;t}\cdot \vec S_{i;b}
    +V\sum_i n_{i;t}n_{i;b} -J_H \sum_i \vec S_{i;l;c}\cdot \vec S_{i;l} 
    \label{eq:1}
    \end{align}
where $c_{i;l;\sigma}$ is an itinerant electron operator of $d_{x^2-y^2} $ orbital with $\vec S_{i;l;c}=\frac{1}{2} \sum_{\sigma \sigma'}c^\dagger_{i;l;\sigma}\vec \sigma_{\sigma \sigma'}c_{i;l;\sigma'}$. $P$ projects out the double occupancy because we assume a large on-site repulsion for the $d_{x^2-y^2}$ orbital. Meanwhile $\vec S_{i;l}$ is the spin-1/2 operator of the localized spin moment from the $d_{z^2}$ orbital. $J_\parallel$ is the intra-layer super-exchange of electrons and not relevant for the physics. $J_\perp$ is the spin coupling along each rung of the local moment. 
It is note worthy that we assume that $J_\parallel=0$ for the localized spin moments and $t_\perp=0$ for the itinerant electron, also resulting in vanishing $J_\perp$ for the $d_{x^2-y^2}$ orbital. We also add a density-density repulsion $V$ between the two layers at each site.

If we insist that the nickelate system is in the $J_H\gg t$ regime, we restricted Hilbert space with the enforced spin-one moment between $\vec S_{i;l;c}$ and $\vec S_{i;l}$. Our Hilbert space is now $2\oplus 3= 5$ dimensional, consisting of two singlon and three doublon states.  
The two singlon states, $\ket{\sigma}=c^{\dag}_{2,\sigma}|G\rangle$, are formed by the one single local spin-1/2 moment of $d_{z^2}$ orbital with $\sigma=\uparrow, \downarrow$. 
Meanwhile, the three spin-triplet doublon states are $\ket{-1}=c^\dagger_{1,\downarrow}c^\dagger_{2,\downarrow}\ket{G}$, $\ket{0}=\frac{1}{\sqrt{2}}(c^\dagger_{1,\uparrow}c^\dagger_{2,\downarrow}+c^\dagger_{1,\downarrow}c^\dagger_{2,\uparrow})\ket{G}$ and $\ket{1}=c^\dagger_{1,\uparrow}c^\dagger_{2,\uparrow}\ket{G}$. 
$\ket{G}$ indicates the empty site, $c^\dagger_{1,\sigma}$ ($c^\dagger_{2,\sigma}$) is labeled for electron operator of $d_{x^{2}-y^{2}}$ ($d_{z^2}$) orbital, omittig the site $i$ and layer index $l$, for simplicity. 
The spin operators for the \textit{spin-1/2} singlon are $ \vec S_{i;l}=\frac{1}{2}\sum_{\sigma \sigma'} \ket{\sigma}_{i,l} \vec \sigma_{\sigma \sigma'}\bra{\sigma'}_{i,l}$ with the Pauli matrices, $\vec \sigma$. 
The spin operators for the \textit{spin-one} doublon states are $ \vec T_{i;l}=\sum_{\alpha,\beta=-1,0,1} \vec T_{\alpha \beta]}\ket{\alpha}_{i,l} \bra{\beta}_{i,l}$, where we have $ T_z=\begin{pmatrix} 1 & 0 & 0 \\ 0 & 0 & 0 \\ 0 & 0 & -1 \end{pmatrix} $, $ T_x=\frac{1}{\sqrt{2}}\begin{pmatrix} 0 & 1 & 0 \\ 1 & 0 & 1 \\ 0 & 1 & 0 \end{pmatrix}$ and $ T_y=\frac{1}{\sqrt{2}}\begin{pmatrix} 0 & -i & 0 \\ i & 0 & -i \\ 0 & i & 0 \end{pmatrix}$ in $\ket{1},\ket{0},\ket{-1}$ basis. The type-II t-J model \cite{oh2023type} is then written as,
\begin{eqnarray}
     H=H_{K}&+&   
     J^{ss}_\perp \sum_{i} \vec s_{i;t}\cdot \vec s_{i;b}+J^{dd}_\perp \sum_i \vec T_{i;t}\cdot \vec T_{i;b}  \notag \\
&+&J^{sd}_\perp 
 \sum_i (\vec S_{i;t}\cdot \vec T_{i;b}+\vec T_{i;t}\cdot \vec S_{i;b})\notag \\
&+&J^{ss}_\parallel \sum_{l, \langle i,j\rangle} \vec s_{i;l}\cdot \vec s_{j;b}
+J^{dd}_\parallel 
\sum_{l, \langle i,j\rangle} \vec T_{i;l}\cdot \vec T_{j;l}  \notag \\
&+&J^{sd}_\parallel
\sum_{l, \langle i,j\rangle} (\vec S_{i;l}\cdot \vec T_{j;l}+\vec T_{i;l}\cdot \vec S_{j;l})
 ,
 \label{type_II_t_J}
\end{eqnarray}
with $J^{ss}_{\perp}=2J^{sd}_{\perp}=4J^{dd}_{\perp}=J_{\perp}$, and 
$J^{ss}_{\parallel}=0$, $J^{sd}_{\parallel}=t_{\parallel}^2/(-2J_{H})$, $J^{dd}_{\parallel}=J_{\parallel}/4$.

\begin{figure}[tb]
    \centering
\includegraphics[width=0.52\textwidth]{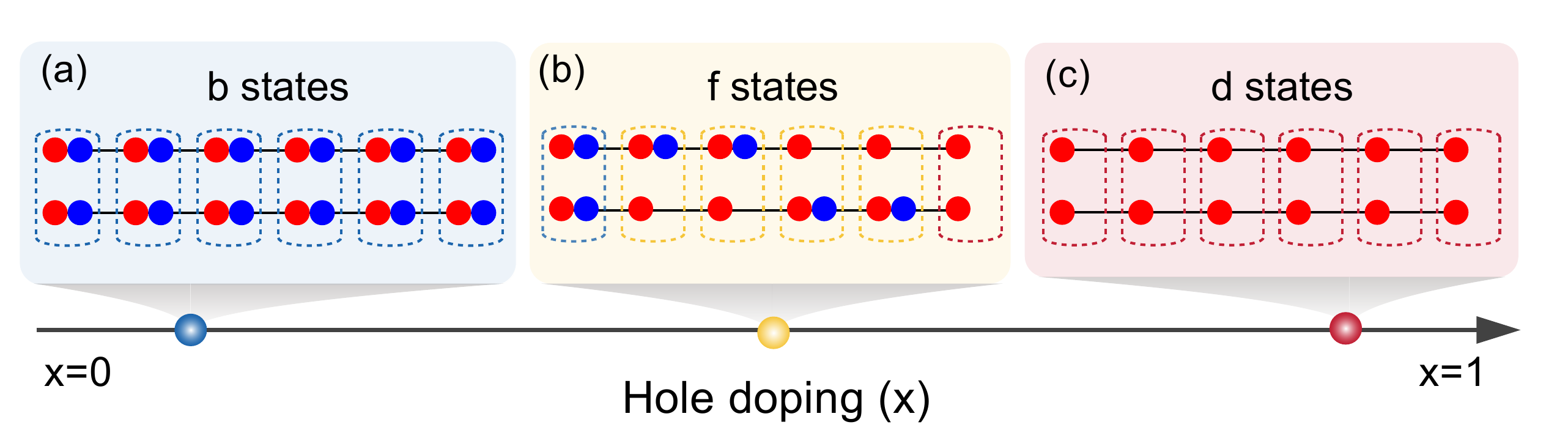}
    \caption{
    \textbf{Schematic illustration of doping dependence of the six states in the ESD t-J model for nickelate.}
    Red and blue circles are the electrons in the $d_{z^2}$ and $d_{x^2-y^2}$ orbitals, respectively. 
    Blue, yellow and red dashed lines denote the $b$, $f$ and $d$ states.  
    (a) At $x=0$, both orbitals are occupied. This is the $b$ state, rung singlet of spin one Mott insulator. (b) At small x, singlon states are doped into the background of $b$ states with one single hole at one rung.  There are also $d$ states with two holes at one rung, especially when $x>0.5$. (c) At $x=1$,
    only the $d_{z^2}$ orbital is occupied and the $d_{x^2-y^2}$ is empty. This is the $d$ state, rung singlet of spin 1/2 Mott insulator.
    }
    \label{fig:filling}
\end{figure}

\subsection{ESD t-J model} \label{secIVA}

Now we show an ESD-t-J model is also responsible for the bilayer type II t-J model in the large $J_\perp$ regime. Again we should keep six states at each rung, with $\ket{b}$, $\ket{d}$ and four singlon states.  The only difference is that now $\ket{d}$ is a rung singlet of spin-1/2 moments and $\ket{b}$ is a rung singlet of spin-one moments. Meanwhile four singlons now should be viewed as polarons formed by hole from the $d_{x^2-y^2}$ orbital hybridized with the localized spin moment from the $d_{z^2}$ orbital. These states are illustrated in Fig.~\ref{fig:filling}.
Here, we leave the detailed formulation in the Appendix, \ref{secB}, 
but the resultant ESD t-J model now becomes, 
\begin{align}
 H&=-t\sum_{l=t,b} \sum_{\langle ij \rangle} 
 Pc^\dagger_{i;l;\sigma} c_{j;l;\sigma}P+\sum_{i}\epsilon
\left(
n_{d;i} +n_{b;i}\right),
\label{ESD_type_II_t_J}
\end{align}
where we have ignored the $J_{\parallel}$ term. 
$\epsilon$ is the average energy off set of the empty state $\ket{b}$ and the doublon $\ket{d}$.
We have $\epsilon=(V-\frac{1}{4}J_\perp)/2$ which is different from the value derived from the one-orbital model, $\epsilon=(V-\frac{3}{4}J_\perp)/2$. 
Employing the generalized slave boson mean-field calculation, we obtain a similar phase diagram with sFL phase and doping induced BCS-BES crossover and pairing dome around $x\simeq 0.5$, indicating their realizations in the Nickelate system (see Appendix Section \ref{secB} for details of the mean-field results ).

\section{Numerical result of the type II t-J model}
\label{secV}
\begin{figure}[b]
    \centering        \includegraphics[width=1.0\linewidth]{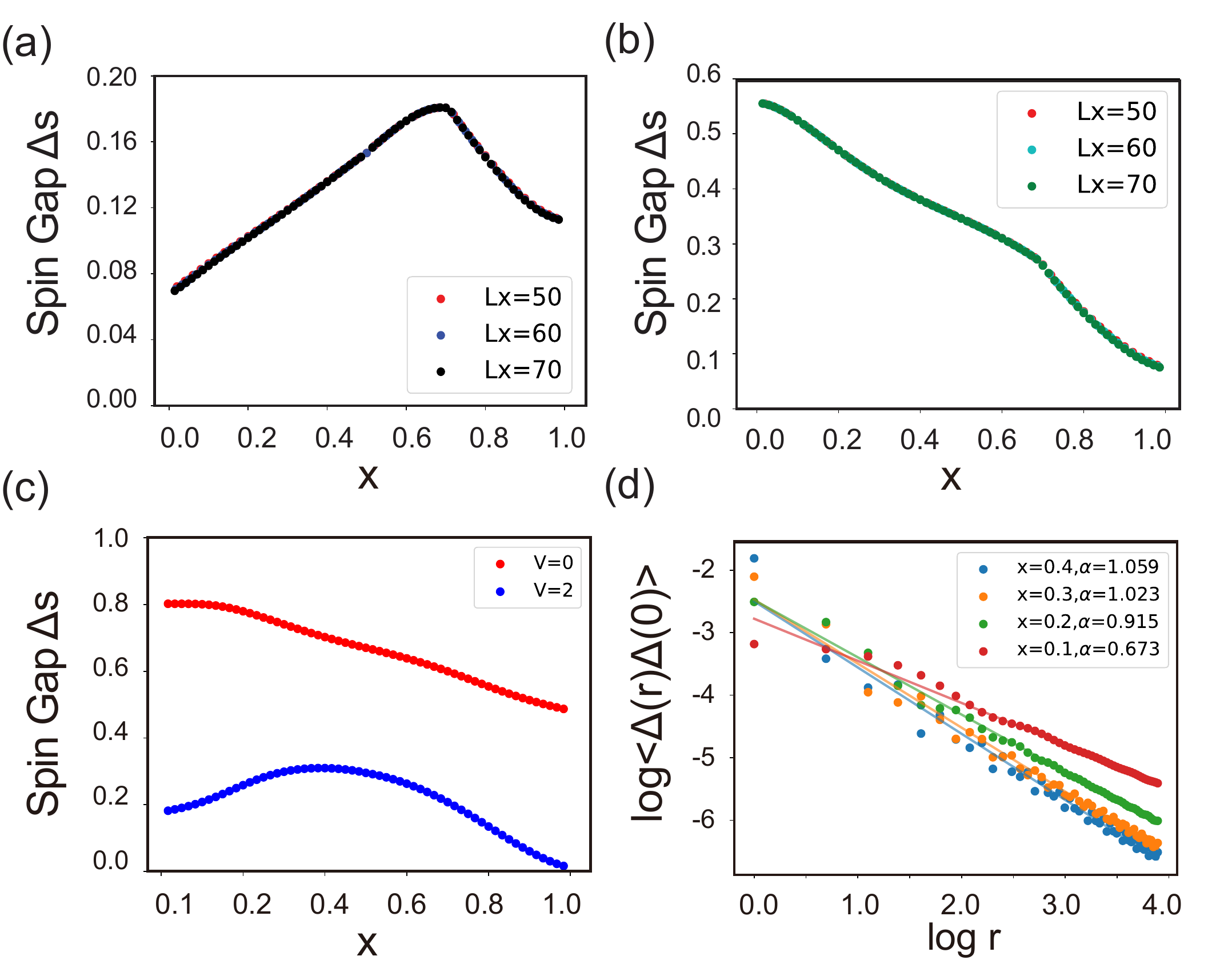}
    \caption{
    \textbf{The DMRG results in one dimension with $L_z=2, L_y=1$.}
    (a) The doping dependence of the
    spin gap $\Delta_S(x)=E(S^z=1)-E(S^z=0)$ of the type-II t-J model at $V=0$. 
    We used $J^{ss}_{\perp}=1$, $J_\parallel^{ss}=0.5$ and the relation $J^{ss}=2J^{sd}=4J^{dd}$ holds for both $J_\parallel$ and $J_\perp$. There is a dome structure for the spin gap, and thus the pairing strength. (b) In contrast, for the one-orbital model with $J_\perp=1$ and $J_\parallel=0.5$, the spin gap monotonically decreases with $x$.  Here in both (a) and (b) we used bond dimension $\chi=2000$, and the results converge for system size $L_x=50,60,70$. (c) Spin gap in the type-II t-J model with $J_{\perp}^{ss}=5$ and $J_\parallel=0$ for $V=0$ and $V=2$.  One can see that the pairing dome disappears at $V=0$ but comes back at $V=2$.  (d) The pairing correlation functions show the power-law decaying behaviour, $\langle\Delta^\dag(r)\Delta(0)\rangle\sim |r|^{-\alpha}$, where the power law exponent $\alpha$ is smaller than one from iDMRG calculation of the type II t-J model at   $J_{\perp}^{ss}=2$ and $J_\parallel=0$ for the dopings $x=0.1,0.2,0.3,0.4$. The results are consistent with a Luther-Emery liquid with power-law superconductivity pairing correlations.
    }
    \label{fig:dmrg_result}
\end{figure}

To support our analytical treatment, we present the numerical results through the density matrix renormalization group (DMRG) simulation \cite{PhysRevLett.69.2863,10.21468/SciPostPhysLectNotes.5} of the type II t-J model, Eq.~\ref{type_II_t_J}, in a two-leg ladder configuration ($L_z=2,L_y=1, L_x \rightarrow +\infty$ ). In our simulation, we  use the parameters $t=1$ ($t_{\perp}=0$) and impose $4J^{dd}_{\perp}=2J^{sd}_{\perp}=J^{ss}_{\perp}$ unless otherwise stated.   While the DMRG algorithm is restricted to the quasi one-dimension, we stress that the pairing in our bilayer model is mediated by an on-site Cooper pair and thus the dimension may not be crucial. Therefore, the essential physics may be already captured by the one-dimensional calculations.  Indeed we will see qualitatively similar results from the DMRG in 1D compared to our slave boson theory of the 2D model.

\subsection{Evidences for Luther-Emery liquid}
In one dimension, true long range order is impossible, thus the superconductor phase we discussed in the previous sections can only manifest itself as a Luther-Emery liquid\cite{PhysRevLett.33.589} with power-law pairing correlations. Indeed we find a finite spin gap in the entire range of the hole doping in the type II t-J model in Eq.~\ref{type_II_t_J} and also in the one-orbital bilayer t-J model in Eq.~\ref{eq:one_orbit_t_J}. As shown in Fig.~\ref{fig:dmrg_result}(a)(b), there is a pairing dome at $J_\perp=1$, $V=0$ in the type II t-J model, but no dome exists in the one-orbital bilayer model at $J_\perp=1,V=0$. We note that at the same value of $J_\perp$, the one-orbital model is closer to the BEC limit. Indeed the type II t-J model has smaller spin gap and is more consistent with BCS picture in the $x\rightarrow 0$ and $x\rightarrow 1$ limit. Hence we find doping induced BCS to BEC crossover as we identified in our analytical theory in Sec.~\ref{secIII}. On the other hand, the one-orbital model seems to be in the BEC limit for the whole doping regime and thus a pairing dome is absent.  Consistent with this picture, we increases $J_\perp$ to $5$ in the type II t-J model and the dome disappears now for $V=0$ and the spin gap is similar to the one-orbital model at $J_\perp=1$ (see Fig.~\ref{fig:dmrg_result}(c)). Next we also add a $V=2$, then the pairing gap is suppressed and the pairing dome comes back. In Fig.~\ref{fig:dmrg_result}(d), we further confirm that there is slow power-law pairing correlations with the power-law exponent smaller than $1$. In the appendix, we show that the central charge is $c=1$, in agreement with a Luther-Emery liquid phase. Overall, our results are qualitative consistent with our analytical analysis of the ESD t-J model in two dimension.

\subsection{Evidence of the small Fermi surface phase}

\begin{figure}[ht]
    \centering        \includegraphics[width=1.02\linewidth]{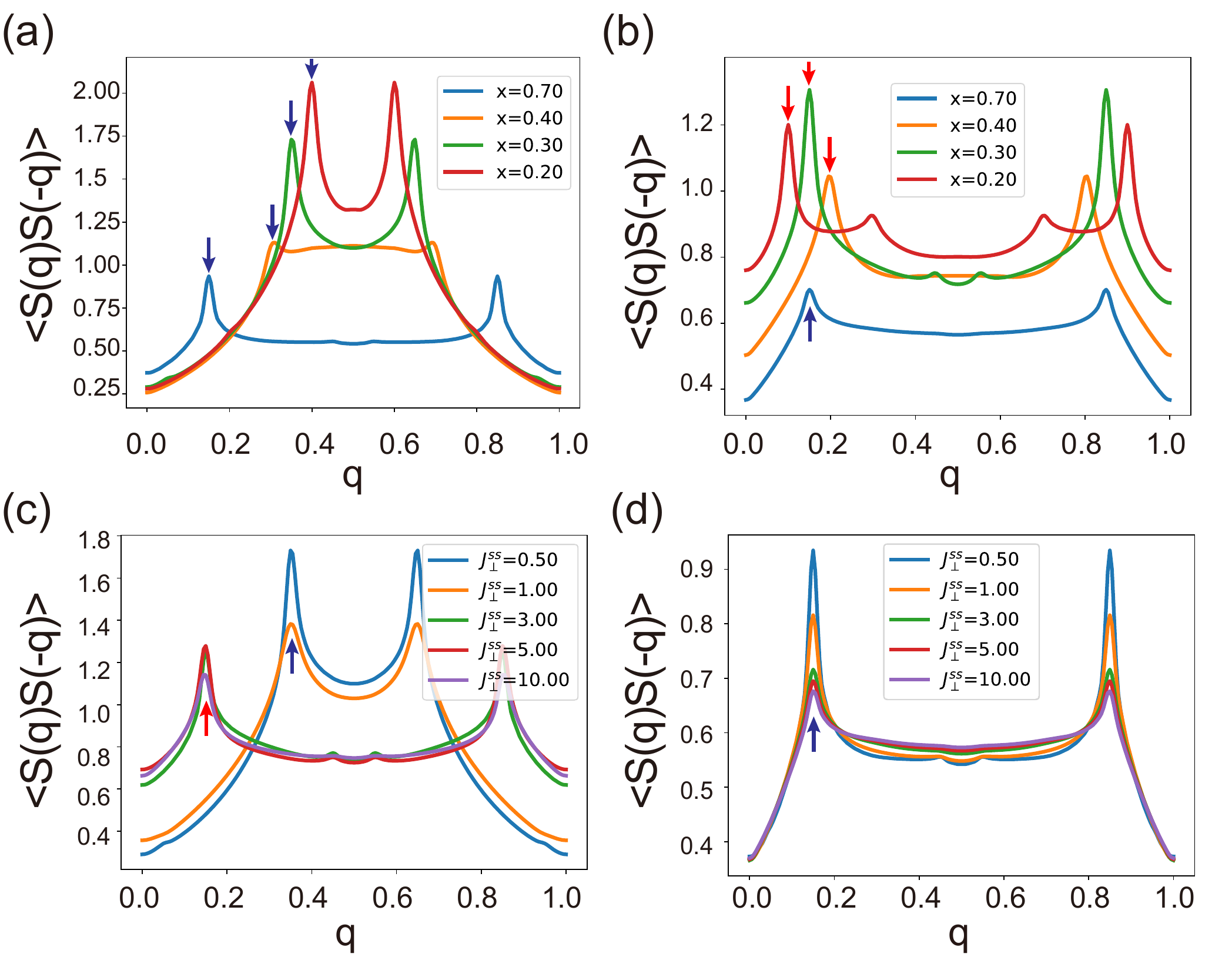}
    \caption{
    \textbf{Numerical evidence of small and large Fermi surfaces with $L_z=2, L_y=1, L_x \rightarrow \infty$.} The iDMRG is performed with the bond dimension $\chi=3000$. We use the parameters  $J_{p}=0.5$, $J_\parallel^{ss}=0$, $J_\parallel^{dd}=0.1$, $J_\parallel^{sd}=0.2$, and $V=100$. The relation of the interlayer coupling is $J_\perp^{ss}=2J_{\perp}^{sd}=4J_\perp^{dd}=J_\perp$. The blue and red arrow indicate $q=\frac{1-x}{2}$ and $\frac{x}{2}$ in units of $2\pi$ respectively. (a) spin-spin correlation in the bottom layer for fixed $J_\perp=0.5$ with the doping $x=0.2,0.3,0.4,0.7$. The peak is located at $q=\frac{1-x}{2}$, consistent with large Fermi surface at small hole doping $x$. (b) spin-spin correlation for fixed $J_\perp=4$ with the doping $x=0.2,0.3,0.4,0.7$, which shows small Fermi surface when $x<0.5$ as illustrated in Fig.~\ref{fig:global_phase_diagram}(b).
    (c) Spin-spin correlation for fixed doping $x=0.3$ with $J_\perp=0.5,1,3,5,10$. There is a transition from large to small Fermi surface when increasing $J_\perp$. (d) Spin-spin correlation for fixed doping $x=0.7$ with the $J_\perp=0.5,1,3,5,10$. The Fermi surface size is always at $q=\frac{1-x}{2}$ as from free fermion model. 
    }
    \label{fig:sFL_FL_phase}
\end{figure}

In addition to superconductor, the other major discovery from our analytical theory in Sec.~\ref{secIII} is the existence of two normal states: the FL phase and the sFL phase with small Fermi surfaces. Here we demonstrate their existences in the 1D model. Strictly speaking, the ground state is always in a superconductor (Luther-Emery liquid in 1D) and the metallic phase is defined only at finite temperature where the pairing is suppressed.  Because DMRG is convenient only for the ground state calculation, here we suppress the pairing through increasing the repulsion $V$.  We will use $V=100$, where the pairing and the spin gap is negligible\footnote{It is quite challenging to determine whether a gap is truly zero or finite but very small.}. Then we can determine the size of the Fermi surface through $2k_F$ singularity in the spin-spin static structure factor $\langle\vec S(q)\cdot \vec S(-q))\rangle$. At the infinite V limit, there is a tendency to freeze the layer polarization $P_{i;z}=\frac{1}{2}(n_{i;t}-n_{i;b})$. So we add a small layer-pseudospin interaction to Eq.~\ref{type_II_t_J}\cite{2023arXiv230501702Y},
\begin{align}
    H^\prime=\frac{J_{p}}{2}(P_x(i)P_y(j)+P_y(i)P_x(j))(4\vec{S}(i)\cdot\vec{S}(j)+1),
\end{align}
in which $\vec{P}$ is the layer pseudospin. The four-flavor spin-layer operator is defined as $P_a S_b=\frac{1}{2}c^\dag_{l\sigma}\tau^a_{ll^\prime}\sigma^b_{\sigma \sigma'}c_{l^\prime\sigma'}$, with $\tau$ as Pauli matrix in the layer space and $\sigma$ as Pauli matrix in the spin space. We add $J_{p}=0.5$ and also $J_\parallel^{ss}=0$, $J_\parallel^{dd}=0.1$, $J_\parallel^{sd}=0.2$ with     $J_{\perp}^{ss}=2J_{\perp}^{sd}=4J_{\perp}^{dd}=J_\perp$ in Eq.~\ref{type_II_t_J}. 
In Fig.~\ref{fig:sFL_FL_phase}(a)(b), for small hole doping $x<0.5$, we find large Fermi surface with $2k_F$ peak at $q=\frac{1-x}{2}$ in units of $2\pi$ for $J_\perp=0.5$, but small Fermi surface with $2k_F$ peak at $q=\frac{x}{2}$ for $J_\perp=4$.  For $x>0.5$ the Fermi surface size is always $\frac{1-x}{2}$ connected to the free fermion result. In Fig.~\ref{fig:sFL_FL_phase}(c)(d) we further confirm that there is a small to large Fermi surface evolution at $x=0.3$ when decreasing $J_\perp$, while the Fermi surface size is fixed for $x=0.7$. For $x=0.3$, in the appendix we show that the momentum distribution function $n(k)$ at large $J_\perp$ indicates a small hole pocket centered at $k=\pi$, which is similar to the sFL phase illustrated for 2D in Fig.~\ref{fig:global_phase_diagram}(b).

While the current calculation uses unrealistic parameter $V=100$ to suppress the pairing, it clearly indicates the existences of the small Fermi surface phase at $x<0.5$ proximate to the Luther-Emery liquid. It is then not hard to imagine the existence of the small Fermi surface phase at finite temperature above the pairing scale at reasonable value of $V$, as we demonstrated in our analytical theory in Sec.~\ref{secIII}. Our discussion within the ESD t-J model in Sec.~\ref{secIII} is restricted to the large $J_\perp$ regime. For $x<0.5$, there is also small to large Fermi surface evolution when decreasing $J_\perp$. The nature of such evolution (or transition through an unusual quantum criticality) is an interesting question given that there is no symmetry breaking order parameter and thus  the conventional Landau-Ginzburg framework is clearly invalid. Especially such an evolution or transition may be realized in the nickelate material where $J_\perp$ can be controlled by pressure.  We leave it to future work to understand the nature of this evolution or transition theoretically.

\section{Summary}\label{secVI}

In summary, we provide one concrete and controlled theory of  an unconventional normal state with small Fermi surface (but no symmetry breaking) and superconductivity at lower temperature. Introducing a new model dubbed as the ESD t-J model defined on a bilayer square lattice, we present a comprehensive phase diagram with the hole doping $x$ using a generalized slave boson theory. Notably, our investigation reveals an unusual dome-shaped structure in the pairing gap, centered around $x=0.5$, in stark contrast to the conventional t-J model, shedding light on novel pairing mechanisms and new physics. On the two sides of the dome, we observe two distinct normal states: the conventional Fermi liquid (FL) phase for $x>0.5$ and an unconventional second Fermi liquid (sFL) for $x<0.5$.  The sFL phase has a Fermi surface volume smaller than half of the Brillouin zone per flavor than the non-interaction limit, which is clearly beyond any weak coupling theory. Furthermore, we identify a doping-induced BCS to BEC crossover, particularly as we tune the doping towards $x=0.5$. Our study also suggests that the bilayer nickelate superconductor La$_3$Ni$_2$O$_7$ should be a promising platform for investigating doping-tuned Feshbach resonances, the emergence of an unconventional symmetric pseudogap metal in the normal state and possible small to large Fermi surface transition without symmetry breaking order parameters.

\textit{Note added}: When finalizing the manuscript, we become aware of a preprint\cite{2023arXiv230913040L}, which also suggests a Feshbach resonance picture based on DMRG simulation on a two-leg ladder of the bilayer one-orbital $t-J_\perp-V$ model. 

\textit{Acknowledgement}: 
This work was supported by the National Science Foundation under Grant No. DMR-2237031.

\bibliographystyle{apsrev4-1}

\bibliography{ref}

\appendix
\onecolumngrid

\section{Slave-boson theory of one-orbital t-J model }\label{secA}
\subsection{Mean-field Hamiltonian and Details on mean-field results}
\label{secA1}
We start with the ESD t-J model Hamiltonian, Eq.~(\ref{esd_one_orbital}) derived in the large $J_{\perp}$ regime, 
\begin{align}
 H&=-t\sum_{l=t,b} \sum_{\langle ij \rangle} c^\dagger_{i;l;\sigma} c_{j;l;\sigma}+H.C.+
\sum_{i}
\left(\epsilon_d
n_{d;i} +\epsilon_b n_{b;i}\right)
\end{align}
with $\epsilon_b=V$ and $\epsilon_d=-\frac{1}{4}J_\perp$.

Upon the mean-field decoupling, one can decouple the Hamiltonian as,
\begin{eqnarray}
    H_{MF} 
    &=&-C_{f} \sum_{l,\langle i,j \rangle} f^{\dagger}_{i;l;\sigma}f_{j;l;\sigma}
    +D_{f} \sum_{\langle i,j \rangle }
\left[
f^{\dagger}_{i;t;\uparrow}f^{\dagger}_{j;b;\downarrow}
-f^{\dagger}_{i;t;\downarrow}f^{\dagger}_{j;b;\uparrow}
\right] +h.c. \notag \\
&+& 
\sum_{i}
[\lambda_{d} d_{i} + H.C.]
+
\sum_{i}
[\lambda_{b} b_{i}+ H.C.]
+\delta_{f} \sum_{i}n_{f;i}
+\delta_{d}\sum_{i}n_{d;i}
+\delta_{b}\sum_{i}n_{b;i}
\label{esd_mft}
\end{eqnarray}
with 
\begin{eqnarray}
    C_{f} &=&t\left[  |\langle d \rangle|^2 -\frac{1}{2} |\langle b \rangle|^2\right], \\
    D_{f} &=& -2 t\frac{1}{\sqrt{2}} \langle d \rangle \langle b\rangle , \notag \\
    \lambda_{d} & = &-2t\left[
   \chi \langle d \rangle^{*} 
    -
    \frac{1}{\sqrt{2}} \Delta^{*}\langle b \rangle   \right],\notag\\
    \lambda_{b} &=& 2t\left[   \frac{1}{2}
    \chi
    \langle b \rangle^{*} 
    +\frac{1}{\sqrt{2}}\Delta^{*}  \langle d \rangle 
    \right]
    ,\notag\\
\delta_{f} &= &-\mu_{0}-\mu,\notag\\
\delta_{d}&= &
\epsilon-\mu_{0}-2\mu,\notag\\
\delta_{b} &= &\epsilon-\mu_{0} \notag 
\end{eqnarray}
where two chemical potentials, $\mu_{0},\mu$ are introduced for the two constraints: $n_{d}+n_{f}+n_{b}=1$, and $n_{f}+2n_{d}=2x$.
For the onsite potential, we can use only one parameter of $\epsilon= (\epsilon_d+ \epsilon_b)/2$ which is an average value of $d$ and $b$, since the difference between them is now just a number, which is irrelevant in this problem, due to the number constraints. 
Therefore, in the main text, we have expanded all discussion controlling only one parameter $\epsilon$, for simplicity.

\begin{figure}[tb]
    \centering
    \includegraphics[width=\textwidth]{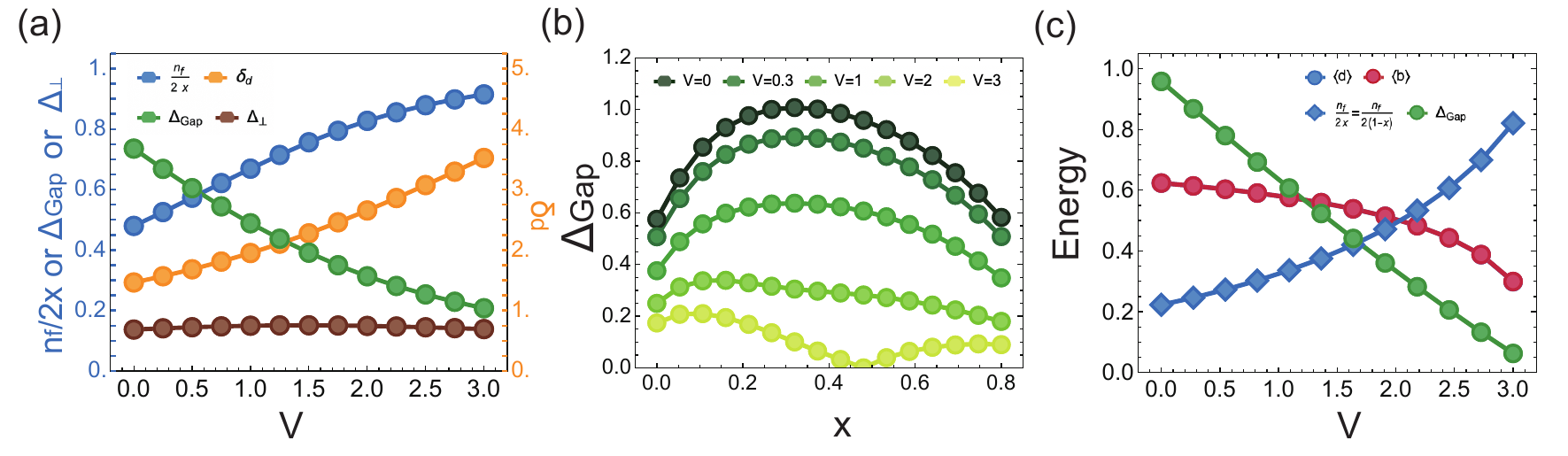}
    \caption{
   \textbf{$V$ dependence of the Slave boson Mean-field results of the one-orbital t-J model.} 
    (a) The single-hole percentage ($\frac{n_f}{2x}$), energy of the virtual pairing ($\delta_{d}$), pairing gap ($\Delta_{\mathrm{Gap}}$), pairing order parameter ($\Delta_{\perp}$) at $x=0.1$. The dependence of $V$ shows that when $V$ is larger, the system moves towards the BCS limit because the energy cost of the virtual Cooper pair ($\delta_d$ for small x) grows linearly with $V$.
    (b) The pairing gap amplitude of various $x$ and $V$ with fixed $J_{\perp}=1$. The variation of color within the plot corresponds to different values of $V=0,0.1,1,2,3$.  The pairing is suppressed by larger $V$ as expected, but there is still a sizable gap $\Delta_{Gap}\sim 0.3 t$ when $V=1$. (c) At $x=0.5$, $\langle d \rangle$ and $\langle b \rangle$ both vanish at larger $V$, in contrast to the behavior at $x=0.1$ in Fig.\ref{fig:mft_part2}. The large $V$ regime of $x=0.5$ is a Mott insulator whose low energy is described by a four-flavor spin model combining the real spin and the layer pseudospin.
 }
\label{fig:mft_part2}
\end{figure}

Then, we consider the four order parameters: $\chi, \Delta, \langle d \rangle, \langle b\rangle$, 
\begin{eqnarray}
    \chi &=& \sum_{\sigma} f_{i;l;\sigma}^{\dagger}f_{j;l;\sigma}, \\
    \Delta  & = & \langle f_{i;t;\uparrow}f_{j;b;\downarrow} -
f_{i;t;\downarrow}f_{j;b;\uparrow} 
     \rangle, \notag\\
    \langle d \rangle &=&\frac{\lambda_d}{-\delta_{d}}, \ \mathrm{for}\ \delta_{d}>0,\notag\\
    \langle b \rangle &=&\frac{\lambda_b}{-\delta_{b}}, \ \mathrm{for}\ \delta_{b}>0,\notag
\end{eqnarray}
By solving the equations self-consistently, we obtained the various observables, as illustrated in 
in Fig.\ref{fig:mft_part1} and \ref{fig:mft_part2}. 
In particular, at $x=0.5$, when we increase $V$, we find that $\langle b \rangle$ and $\langle d \rangle$ both go to $0$, as shown in Fig.~\ref{fig:mft_part2} (c).  This is a Mott insulator where the low energy physics is described by a four-flavor spin model combining the real spin and the layer, similar to what is discussed in Ref.~\onlinecite{zhang2021SU(4)}. The four $f$ fermions are now the famous Abrikosov fermions.  The exact nature of the spin-layer physics of this Mott insulator is not our interest in this paper. In realistic system $V$ is likely not large enough to reach the Mott insulator.

\subsection{Estimation of $T_c$ of BCS' and BCS regime from the fermion-boson model} \label{secA2}
We consider the BCS' limit ($x<0.5$), where the parent state is sFL. 
As discussed in Sec.\ref{secIIIB}, in this limit, the effective model can be described by fermion-boson model Hamiltonian ($H=H_{f}+H_{d}+H_{fd}$), 
\begin{align}
H_{f}&=\sum_{\langle i,j\rangle}t_{\mathrm{eff}}
f_{i,l,\sigma}^\dag f_{j,l,\sigma}
+h.c. -\mu_{f}\sum_{i}n_{f,i,l}, \quad 
H_{d}=\sum_{i}\delta_d n_{d,i} \notag\\
H_{fd}&= \sum_{\langle i, j\rangle}g(d_i^\dag+d_j^\dag)(f_{i,b,\downarrow}f_{j,t,\uparrow}-f_{i,b,\uparrow}f_{j,t,\downarrow})+h.c.\notag
\\
&= 4g\sum_{\vec k,\vec q}
F(\vec k,\vec q)
d_{\vec q}^{\dagger} [
f_{\vec k,b,\downarrow}f_{-\vec k+\vec q,t,\uparrow}
-f_{\vec k,b,\uparrow}f_{-\vec k+\vec q,t,\downarrow}]+h.c.\notag
\end{align}
We have $t_{\mathrm{eff}}=\frac{1}{2}t\langle b\rangle^2$, $g=\frac{1}{\sqrt{2}}t\langle b\rangle$ 
and the vertex function is given by 
\begin{align}
    F(\vec k,\vec q)&=\frac{1}{2}\left[\cos k_x+\cos k_y+\cos(-k_x+q_x)+\cos(-k_y+q_y)\right].\notag
\end{align}
The effective four-fermion interaction can be obtained by integrating the bosonic field, 
\begin{eqnarray}
H_{\mathrm{eff}}&=&
\sum_{k,k',q} V_{\mathrm{eff}}(k,k',q)
 [f_{k,b,\downarrow}f_{-k+q,t,\uparrow}
-f_{k,b,\uparrow}f_{-k+q,t,\downarrow}]
 [f_{-k'+q,t,\uparrow}^\dag f_{k',b,\downarrow}^\dag
-f_{-k'+q,t,\downarrow}^\dag f_{k',b,\uparrow}^\dag]\notag,\\
V_{\mathrm{eff}}&=& 16 g^2 F(k,q)F^*(k',q) G_{b}(q,i\omega_n\simeq 0),
\end{eqnarray}
where $G_{b}^{-1}(q,i\omega_n\simeq 0)=-\delta_{d}$.
Projecting the four-fermion interaction into $q=0$ subspace, where the pairing with total zero momentum lives, the attractive and separable interaction form is achieved,
\begin{eqnarray}
    H_{\mathrm{eff}} &\rightarrow & -v_{\mathrm{eff}}
    \sum_{k,k'} F(k,0)F^*(k',0)
     [f_{k,b,\downarrow}f_{-k,t,\uparrow}
-f_{k,b,\uparrow}f_{-k,t,\downarrow}]
 [f_{-k',t,\uparrow}^\dag f_{k',b,\downarrow}^\dag
-f_{-k',t,\downarrow}^\dag f_{k',b,\uparrow}^\dag]\label{eq:7}.
\end{eqnarray}with strength $v_{\mathrm{eff}}=16\frac{g^2}{\delta_d}$.
Then, the effective fermionic model becomes ($H_{f}
= H_{0}+H_{int}$),
\begin{eqnarray}
H_{0}&=&\sum_{\vec k,\sigma} \epsilon_{\vec k}
f_{\vec k,l,\sigma}^\dag f_{\vec k,l,\sigma}= t_{\mathrm{eff}}\sum_{\vec k,\sigma} (\cos k_x + \cos k_y) f_{k,l,\sigma}^\dag f_{\vec k,l,\sigma},
\notag
\\
H_{int}&=&-v_{\mathrm{eff}}
    \sum_{\vec k} F(\vec k,0)
     [f_{\vec k,b,\downarrow}f_{-\vec k,t,\uparrow}
-f_{\vec k,b,\uparrow}f_{-\vec k,t,\downarrow}]
 \sum_{\vec k'} F^{*}(\vec k',0) [f_{-\vec k',t,\uparrow}^\dag f_{\vec k',b,\downarrow}^\dag
-f_{-\vec k',t,\downarrow}^\dag f_{\vec k',b,\uparrow}^\dag],\notag
\end{eqnarray}
The mean-field approximation with the ansatz
\begin{eqnarray}
  \Delta &=& v_{\mathrm{eff}}\left \langle\sum_{\vec{k}'} F(\vec{k}',0)
     [f_{\vec k',b,\downarrow}f_{-\vec k',t,\uparrow}
-f_{\vec{k}',b,\uparrow}f_{-\vec{k}',t,\downarrow}]\right\rangle,
\end{eqnarray}
gives the mean-field Hamiltonian,
\begin{eqnarray}
H_{MF}
&=&
\sum_{\vec k,\sigma} \epsilon_{\vec k}f_{\vec k,l,\sigma}^\dag f_{\vec k,l,\sigma}-
 \sum_{\vec k}   \Delta F(\vec k',0) [f_{-\vec k,t,\uparrow}^\dag f_{\vec k,b,\downarrow}^\dag
-f_{-\vec k,t,\downarrow}^\dag f_{\vec k,b,\uparrow}^\dag]+h.c.+\frac{\Delta^2}{v_{\mathrm{eff}}}.\notag
\end{eqnarray}
The mean-field free energy is
\begin{eqnarray}
    F_{MF}&=& 
    -2T \sum_{\vec k}\log \left( 2 \cos \frac{E_{\vec k}}{2T}\right)+\frac{\Delta^2}{v_{\mathrm{eff}}},\notag
 \end{eqnarray}
 with $E_{\vec k}=[\epsilon^{2}_{\vec k} +\Delta^2 (\cos k_x +\cos k_y)^2]$.
Differentiating the free energy and taking the limit $\Delta\rightarrow 0$ gives the critical temperature $T_c$,
\begin{eqnarray}
     \frac{1}{v_{\mathrm{eff}}}
     =\sum_{\vec k}\tanh\frac{\epsilon_{\vec k}}{2T_c}\frac{ (\cos k_x +\cos k_y)^2}{\epsilon_{\vec k}}. \notag
\end{eqnarray}
We can replace the summation over momentum, $\sum_{\vec{k}}$, with the integration over the energy $\int d\epsilon D(\epsilon)$, with the density of state of the normal state, $D(\epsilon)$, 
\begin{eqnarray}
     \frac{1}{v_{\mathrm{eff}}}
     &=&
     \int^{W}_{0} d\epsilon D(\epsilon) \tanh\frac{\epsilon_{k}}{2T_c}\frac{ (\cos k_x +\cos k_y)^2}{\epsilon_{k}}\notag
     \\
     &=&  D(0)C_{FS}
     \int^{W}_{0} d\epsilon \frac{\tanh\frac{\epsilon}{2T_c}}{\epsilon}\notag\\
     &=& \log\left( \frac{W}{2T_c}\right)-\log \frac{\pi}{4} +\gamma. \notag
\end{eqnarray}
Here, we introduce the UV energy cutoff as a band-width, $W$. In the second equality, we approximated that the dominant contribution in $\sum_{\vec{k}}$ is from near the zero energy. 
$D_{0}$ is the zero-energy density of state, and  $C_{\mathrm{FS}} =\langle (\cos k_x +\cos k_y )^2\rangle_{\mathrm{FS}}$ is an average of the form factor over the normal Fermi surface.  
Apparently, $C_{\mathrm{FS}}$ encodes the microscopic details of the system, such as the shape of the Fermi surface. 
The final equality holds for the limit $\Lambda \gg T_{c}$, where the Euler constant $\gamma=0.577$ is used.
Finally, the critical temperature of BCS' limit becomes 
\begin{eqnarray}
    T_{c}
    =
    \frac{2}{\pi}e^{\gamma}W\exp\left[ -\frac{\delta_{d}}{16g^2 D(0)C_{\mathrm{FS}}}\right],
\end{eqnarray}
with $v_{\mathrm{eff}}=16\frac{g^2}{\delta_{d}}$ and $g=\frac{t\langle b \rangle}{2}$. 
Similarly, the critical temperature of BCS regime is,
\begin{eqnarray}
    T_{c}
    =
    \frac{2}{\pi}e^{\gamma}W\exp\left[ -\frac{\delta_{b}}{16g^2 D(0)C_{\mathrm{FS}}}\right].
\end{eqnarray}
But here we have $t_{\mathrm{eff}}=t \langle d \rangle^2$, $v_{\mathrm{eff}}=16\frac{g^2}{\delta_{b}}$ and $g=t\langle bd \rangle$.

\begin{figure}[h]
    \centering
\includegraphics[scale=0.7
     ]{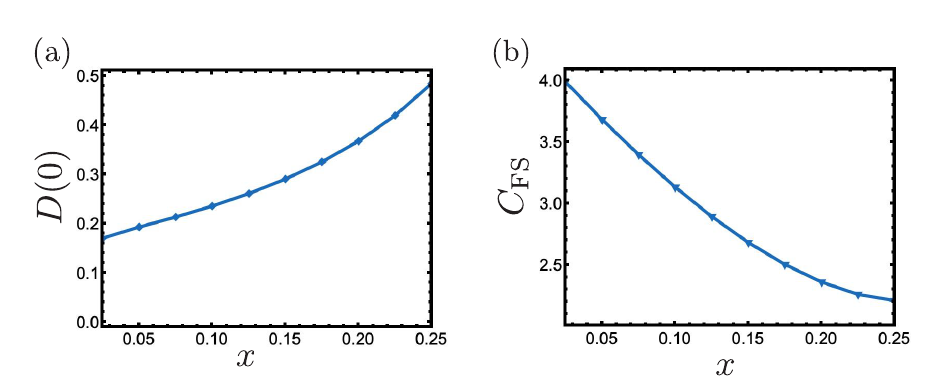}
    \caption{Hole doping dependence of (a) zero-energy density of states $D(0)$, (b) $C_{\mathrm{FS}}=\langle (\cos k_x + \cos k_y)^2\rangle_{\mathrm{FS}}$ 
These phenomenological values, $D(0), C_{\mathrm{FS}}$, determine the critical temperature near the BCS limit (near $x\simeq 0$).
     }
    \label{fig:dos_dep}
\end{figure}

\newpage

\section{Slave-boson theory of type-II t-J model }
\label{secB}
\subsection{Exact form of the six states} 
\begin{figure}[htbp]
    \centering

\includegraphics[width=0.9\textwidth]{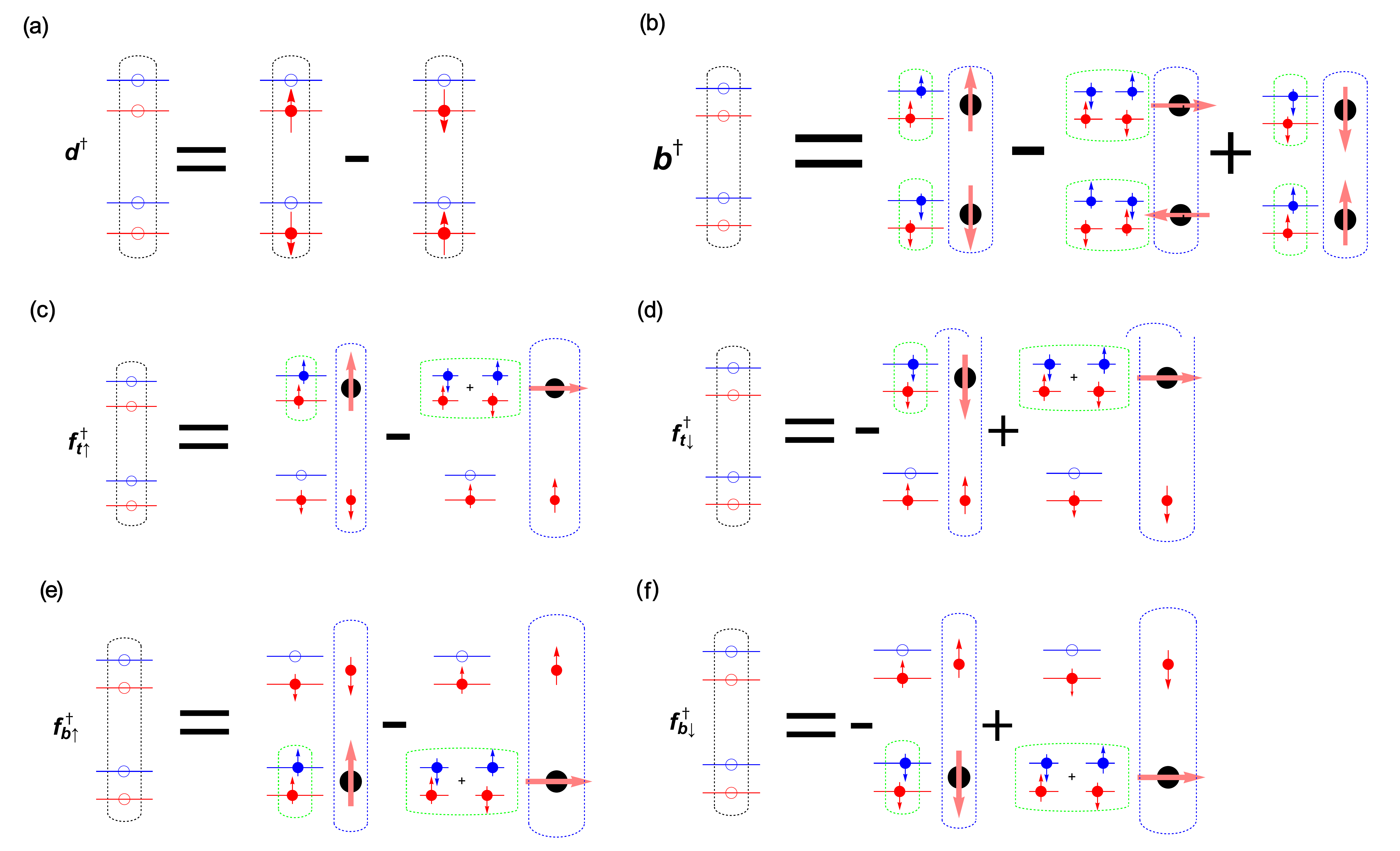}
    \caption{Illustrated six states in the ESD-t-J model. The empty circles correspond to holes in the $d_{z^2}$ orbital(red), and $d_{x^2-y^2}$ orbital (blue), while the solid circle corresponds to filled electrons in that orbital. 
    The red and blue lines correspond to spin 1/2 moment of the  $d_{z^2}$ and $d_{x^2-y^2}$ orbitals. 
    The black circle labels the spin-one state formed by the electrons in $d_{x^2-y^2}$ orbit and $d_{z^2}$ orbit.  $b^\dagger \ket{0}$ and $d^\dagger \ket{0}$ create rung-singlet of spin-one and spin-half moments, respectively. $f^\dagger_{a\sigma}$ creates a state with one hole and a total $S=1/2$ and can be viewed as a polaron. We ignored the coefficients of the superposition, for simplicity, and see Eq.~\ref{eq:parton} for the details.}
    \label{fig:partons_type_II}
\end{figure}

In this section, we provide the detailed derivation of the ESD model starting from the type-II t-J model.
We remark that our main focus is the in $t\ll J_\perp\ll J_{H}$ limit.
In the $J_{H}\gg J_{\perp}$ case, if both two orbits are occupied with electrons, they should form the triplet state carrying the total spin-one.  
Then, we can consider the $J_\perp$ term, only keeping the $6$ states per site, combining two layers together. These $6$ states as illustrated in Fig.~\ref{fig:partons_type_II} can be specified by the number of holes ($n_h^T$) and the total spin ($S_{tot}$), as follows. 
\begin{itemize}
    \item ${d}$ ($n_h^T=2$, $S_{tot}=0$): A rung singlet of two spin-half moments of $d_{z^2}$ with the $d_{x^2-y^2}$ orbital empty.  
    \item $b$ ($n^T_h=0$, $S_{tot}=0$): A rung singlet between two spin-one moments (hybridizing of both $d_{x^2-y^2}$ and $d_{z^2}$). 
    \item $f_{l,\sigma}$ ($n^T_h=1$, $S_{tot}=\frac{1}{2}$): Four polaron states labeled by  $l=t,b$ and $\sigma=\uparrow,\downarrow$ carrying total 1/2 spin moment. 

\end{itemize}

 In terms of the microscopic electron operators $c_1 (c_2)$ for $d_{x^2-y^2}$ ($d_{z^2}$) orbital, these six states can be written as 
\begin{align}
    \ket{d}=& \frac{1}{\sqrt{2}}[c_{b;2;\downarrow}^\dag c_{t;2;\uparrow}^\dag-c_{b;2;\uparrow}^\dag c_{t;2;\downarrow}^\dag]|G\rangle
=\frac{1}{\sqrt{2}}[|S_{t}=\frac{1}{2},S_{b}=-\frac{1}{2}\rangle -|S_{t}=-\frac{1}{2},S_{b}=\frac{1}{2}\rangle],\label{eq:parton}
\end{align}
\begin{align}
    \ket{b} =& \frac{1}{\sqrt{3}}
[ c_{b;2;\downarrow}^\dag
c_{b;1;\downarrow}^\dag 
c_{t;2;\uparrow}^\dag
c_{t;1;\uparrow}^\dag 
  -
  \frac{( c_{b;2;\downarrow}^\dag c_{b;1;\uparrow}^\dag + c_{d;2;\uparrow}^\dag c_{d;1;\downarrow}^\dag)}{\sqrt{2}}
  \frac{(c_{t;2;\downarrow}^\dag c_{t;1;\uparrow}^\dag 
+c_{t;2;\uparrow}^\dag c_{t;1;\downarrow}^\dag )}{\sqrt{2}}  
+c_{b;2;\uparrow}^\dag c_{b;1;\uparrow}^\dag 
c_{t;2;\downarrow}^\dag  c_{t;1;\downarrow}^\dag 
    ]|G\rangle,\notag
    \\
    =& \frac{1}{\sqrt{3}}[|T_{t}=1,T_{b}=-1\rangle -|T_{t}=0,T_{b}=0\rangle +|T_{t}=-1,T_{b}=1\rangle],\notag
    \end{align}
    \begin{align}
\ket{t,\uparrow}
=&\frac{1}{\sqrt{3}}
    [\sqrt{2} 
    c_{b;2;\downarrow}^\dag
    c_{t;2;\uparrow}^\dag
    c_{t;1;\uparrow}^\dag
    - c_{b;2;\uparrow}^\dag
   \frac{ (c_{t;2;\downarrow}^\dag c_{t;1;\uparrow}^\dag
    +c_{t;2;\uparrow}^\dag c_{t;1;\downarrow}^\dag)}{\sqrt{2}}
    ]|G\rangle\notag
    =\frac{\sqrt{2}}{\sqrt{3}}|T_{t}=1,S_{b}=-\frac{1}{2}\rangle -\frac{1}{\sqrt{3}}|T_{t}=0,S_{b}=\frac{1}{2}\rangle,
    \notag\\
\ket{t,\downarrow} =&
\frac{1}{\sqrt{3}}
    [-\sqrt{2} 
    c_{b;2;\uparrow}^\dag
    c_{t;2;\downarrow}^\dag
    c_{t;1;\downarrow}^\dag
    + c_{b;2;\downarrow}^\dag
   \frac{ (c_{t;2;\uparrow}^\dag c_{t;1;\downarrow}^\dag
    +c_{t;2;\downarrow}^\dag c_{t;1;\uparrow}^\dag)}{\sqrt{2}}
    ]|G\rangle =-\frac{\sqrt{2}}{\sqrt{3}}|T_{t}=-1,S_{b}=\frac{1}{2}\rangle +\frac{1}{\sqrt{3}}|T_{t}=0,S_{b}=-\frac{1}{2}\rangle,\notag\\
\ket{b,\uparrow} =&\frac{1}{\sqrt{3}}
    [\sqrt{2} 
    c_{b;2;\uparrow}^\dag c_{b;1;\uparrow}^\dag c_{t;2;\downarrow}^\dag
    - \frac{(c_{b;2;\downarrow}^\dag c_{b;1;\uparrow}^\dag +c_{b;2;\uparrow}^\dag c_{b;1;\downarrow}^\dag)}{\sqrt{2}}c_{t;2;\uparrow}^\dag
    ]|G\rangle= \frac{\sqrt{2}}{\sqrt{3}}|T_{b}=1,S_{t}=-\frac{1}{2}\rangle -\frac{1}{\sqrt{3}}|T_{b}=0,S_{t}=\frac{1}{2}\rangle,\notag\\
\ket{b,\downarrow} =&\frac{1}{\sqrt{3}}
    [-\sqrt{2} 
  c_{b;2;\downarrow}^\dag c_{b;1;\downarrow}^\dag c_{t;2;\uparrow}^\dag
+
(\frac{c_{b;2;\uparrow}^\dag c_{b;1;\downarrow}^\dag +c_{b;2;\downarrow}^\dag c_{b;1;\uparrow}^\dag }{\sqrt{2}})c_{t;2;\downarrow}^\dag 
    ]|G\rangle=-\frac{\sqrt{2}}{\sqrt{3}}|T_{b}=-1,S_{t}=\frac{1}{2}\rangle +\frac{1}{\sqrt{3}}|T_{b}=0,S_{t}=-\frac{1}{2}\rangle\notag.
\end{align}
Here, $\ket{G}$ is the empty state without any electron in neither orbital at each rung. $\ket{G}$ does not belong to our final Hilbert space. $|T_{l}\rangle$ denotes the spin index of spin-1 operator, while $|S_{l}\rangle$ is for the spin-1/2 operator at $l$ layer.
\subsection{Parton construction and the derivation of the ESD model}
To analyze the ESD-t-J model, we use the following slave-boson construction,
\begin{eqnarray}
c_{i;l,\sigma}&=&\frac{\sqrt{3}}{2}d^\dag_{i} f_{i;l,\sigma}+\frac{\epsilon_{\sigma,\sigma'}}{\sqrt{2}}
f_{i,\overline{l},\sigma'}^\dag b_i ,
\label{parton_typeII}
\end{eqnarray}
with introducing the antisymmetric tensor 
and $\epsilon_{\uparrow \downarrow}=-\epsilon_{\downarrow \uparrow}=1$, 
and the opposite layer index $\overline{l}$.
Substituting the parton to the ESD-t-J model, we rewrite it in the form,
\begin{eqnarray}
   H \!\!&=&\!\!-t\sum_{l,\langle i,j\rangle}\!\!
\left[\frac{\sqrt{3}f^\dag_{i,l,\uparrow}d_{i}}{2} +\frac{b_i^\dag f_{i,\overline{l},\downarrow}}{\sqrt{2}}
\right]\!\!
\left[\frac{\sqrt{3}d^\dag_{j} f_{j;l,\uparrow}}{2}+\frac{f_{j,\overline{l},\downarrow}^\dag b_{j}}{\sqrt{2}}
\right]\notag+
\left[\frac{\sqrt{3}f^\dag_{i,l,\downarrow}d_{i}}{2} -\frac{b_i^\dag f_{i,\overline{l},\uparrow}}{\sqrt{2}}
\right]
\left[\frac{\sqrt{3}d^\dag_{j} f_{j;l,\downarrow}}{2}-\frac{f_{j,\overline{l},\uparrow}^\dag b_{j}}{\sqrt{2}}
\right]\notag\\
&&+\sum_{i}
\delta_{f} n_{f;i}
+\delta_{d} n_{d;i}
+\delta_{b} n_{b;i}.
\end{eqnarray}
with $\delta_{f}=-(\mu_{0}+\mu)$, 
$\delta_{d}=\epsilon-(\mu_{0}+2\mu)$, $\delta_{b}=\epsilon-\mu_{0}$, and $\epsilon=(V-\frac{1}{4}J_\perp)/2$.

\subsection{Self consistency equation and Details on mean-field results}
\begin{figure}[tb]
    \centering
    \includegraphics[width=.62\textwidth]{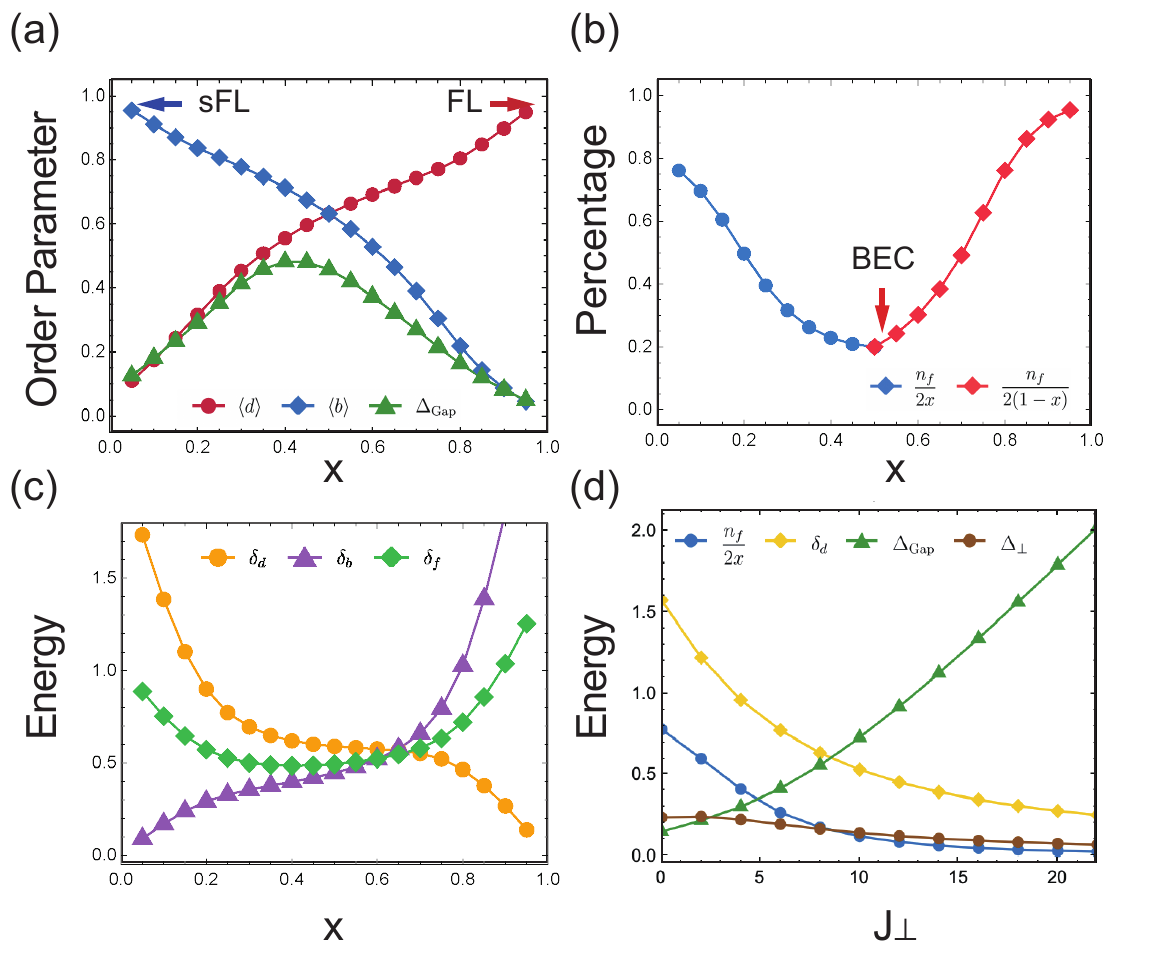}
    \caption{ \textbf{Slave boson Mean-field results of the typeII t-J model.} (a) Doping dependence of order parameters $\langle d\rangle, \langle b \rangle$ and energy gap $\Delta_{\mathrm{Gap}}$ at $J_{\perp}=1$, $V=0.3$.
 Above a small temperature scale, $\langle b \rangle \neq 0, \langle d \rangle=0$ at $x$ close to 0 and $\langle d \rangle \neq 0, \langle b \rangle =0$ close to $x=1$, leading to the sFL phase and the FL phase in the two sides respectively. 
The optimal pairing gap appears near $x\simeq 0.5$. 
(b) The percentage of single hole state $\frac{n_{f}}{2x} (\frac{n_{f}}{2(1-x)})$ at $x<0.5$ ($x>0.5$). 
The single hole percentage is minimized near $x=0.5$, suggesting that the density of a single hole (single electron) is depleted by Cooper pair, a signature of the BEC limit with tightly bound Copper pair as the main carrier.
(c) The value of $\delta_{f},\delta_{d},\delta_{b}$ for showing the energy cost of the virtual cooper pair. 
(d) The single-hole percentage ($\frac{n_f}{2x}$), gap of the virtual Cooper pair ($\delta_{d}$), the order parameter ($\Delta_{\perp}\equiv 
\langle f_{i;t;\uparrow}f_{j;b;\downarrow} -f_{i;t;\downarrow}f_{j;b;\uparrow} 
     \rangle$) and the single particle gap $\Delta_{\mathrm{Gap}}$ at $x=0.1$. The dependence of $J_{\perp}$ shows that when $J_{\perp}$ is larger, the system is moving towards the BEC limit because the virtual Cooper pair energy $\delta_d$ decreases and the singly occupied state $n_f$ is depleted.
     }
    \label{fig:mft_part1_typeII}
\end{figure}

\begin{figure}[tb]
    \centering
    \includegraphics[width=\textwidth]{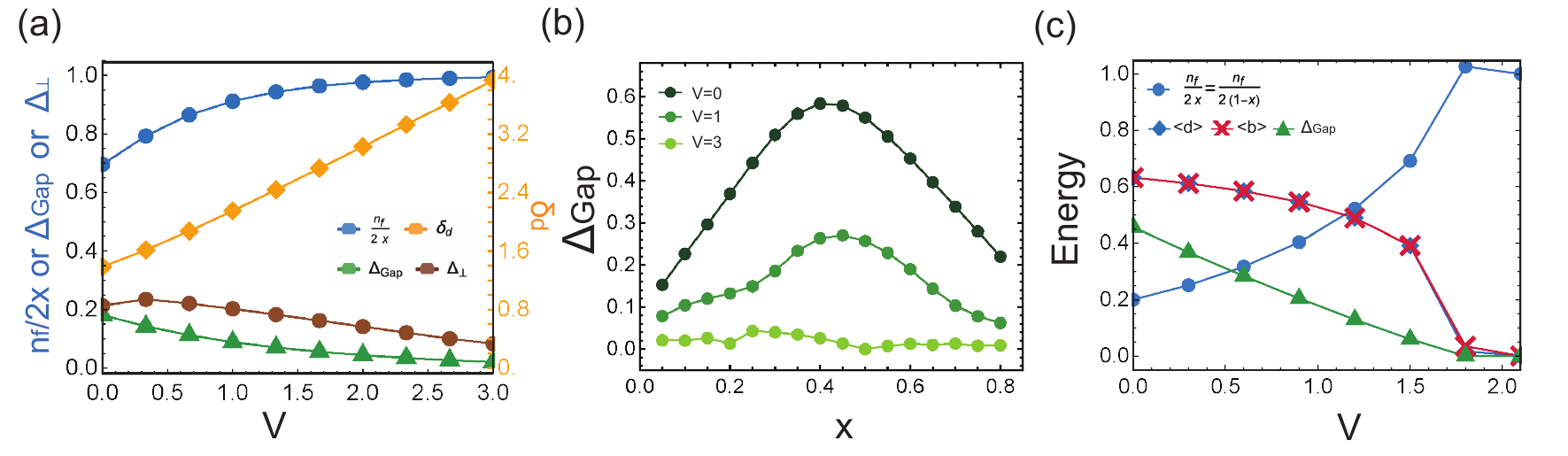}
    \caption{ \textbf{Slave boson Mean-field results of the type-II t-J model.} 
      \textbf{$V$ dependence of the Slave boson Mean-field results of the one-orbital t-J model.} 
    (a) The single-hole percentage ($\frac{n_f}{2x}$), energy of the virtual pairing ($\delta_{d}$), pairing gap ($\Delta_{\mathrm{Gap}}$), pairing order parameter ($\Delta_{\perp}$) at $x=0.1$. The dependence of $V$ shows that when $V$ is larger, the system moves towards the BCS limit.
    (b) The pairing gap amplitude of various $x$ and $V$ with fixed $J_{\perp}=1$. The variation of color within the plot corresponds to different values of $V=0,1,3$.  The pairing is suppressed by larger $V$ as expected (c) At $x=0.5$, $\langle d \rangle$ and $\langle b \rangle$ both vanish at larger $V$, in contrast to the behavior at $x=0.1$ in Fig.\ref{fig:mft_part2}. The large $V$ regime of $x=0.5$ is a Mott insulator.
    }
    \label{fig:mft_part2_typeII}
\end{figure}

For performing the mean-field theory, we introduce the order parameters, $\chi_{l;i,j} \equiv \sum_{\sigma} f_{i;l;\sigma}^{\dagger}f_{j;l;\sigma} $, $\Delta_{i,j}\equiv 
\langle f_{i;t;\uparrow}f_{j;b;\downarrow} -f_{i;t;\uparrow}f_{j;b;\downarrow} 
     \rangle$,$\langle d \rangle$, and $\langle b \rangle$, and the resultant mean-field Hamiltonian is exactly same with that of the one-orbital ESD Mean-field Hamiltonian, Eq.(~\ref{esd_mft}).
However, the mean-field parameters are different because of the difference of coefficients of the two different parton ansatzs (See Eq.(\ref{parton_one}) and Eq.(\ref{parton_typeII})).
The $C_{f},D_{f},\lambda_d,\lambda_b$'s are now 
\begin{eqnarray}
    C_{f} &=&t\left[ \frac{3}{4} |\langle d \rangle|^2 -\frac{1}{2} |\langle b \rangle|^2\right], \\
    D_{f} &=& -2 t\sqrt{\frac{3}{8}} \langle d \rangle \langle b\rangle , \notag \\
    \lambda_{d} & = &-2t\left[   \frac{3}{4} \chi\langle d \rangle^{*}
    -\sqrt{\frac{3}{8}} \Delta^{*}\langle b \rangle 
    \right],\notag\\
    \lambda_{b} &=&2t\left[  \frac{1}{2}\chi\langle b \rangle^{*} 
    +\sqrt{\frac{3}{8}} \Delta^{*} \langle d \rangle 
    \right]
    ,\notag\\
\delta_{f} &= &-\mu_{0}-\mu,\notag\\
\delta_{d}&= &
\epsilon_{b}-\mu_{0}-2\mu,\notag\\
\delta_{b} &= &\epsilon_{d}-\mu_{0}. \notag 
\label{esd_mft_pars}
\end{eqnarray}
The Mean-field results solving the self-consistency equations with Eq.\ref{esd_mft_pars} are presented in Fig.\ref{fig:mft_part1_typeII}, \ref{fig:mft_part2_typeII}. 

\subsection{Experimental implication of BCS to BEC crossover}
\begin{figure}[tb]
    \centering
    \includegraphics[width=\textwidth]{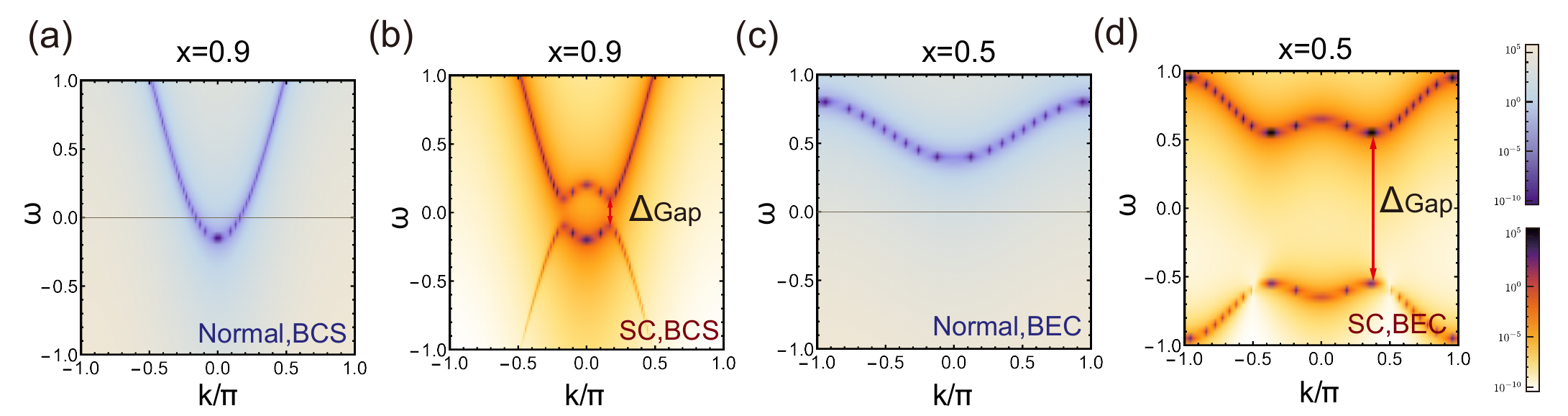}
    \caption{ \textbf{Spectral functions of the normal and superconducting state based on ESD t-J model}
    The parameter is set as $J_{\perp}=5, V=1$. The momentum is along $(-\pi,-\pi)$ to $(\pi,\pi)$.
Two different $x$ values are chosen for illustration to contrast the BCS 
 ($x=0.9$) and BEC ($x=0.5$) limit. 
 The normal band touches the zero energy for a BCS limit (a), while it does not in the BEC limit (c). Here, the normal spectrum is defined as setting $\Delta_{\perp}=0$ from the associated superconducting states. 
 In the superconducting states, the pairing gap, $\Delta_{\mathrm{gap}}$, is determined by two competing characteristic energy scales, chemical potential, $\delta_f$, and the pairing order parameter, $\Delta_{\perp}$. The pairing gap is dominated by $\Delta_{\perp}$ for BCS limit (b), while it is governed by $\delta_{f}$ for BEC limit (d). Notably, the location of the minimum gap shifts away from $k=(0,0)$, even in the BEC limit. This is because the gap function of this system is $s^\prime$-wave with $\Delta_\perp(k)\sim \cos k_x + \cos k_y$, which attains its maximum at $k=(0,0)$. 
}
    \label{fig:spectral}
\end{figure}

In this section, we provide the experimental implication to see the BCS and BEC crossover physics. The BCS-BEC crossover is extensively studied in the context of cold atom experiments \cite{zwerger2011bcs}. Recently, it has been observed in electronic systems such as doped iron-based materials \cite{BEC-BCS1}. 
One of the powerful tools is the Angle resolved photoemission spectroscopy (ARPES) technique by resolving the Bogoliubov quasiparticles dispersions. The cross-over condition is known to be  $\Delta/\epsilon_{F}\sim1$. 
This signifies that in the BCS limit ($\Delta/\epsilon_{F}\ll 1$), the quasiparticle spectrum is dominated by the Fermi energy, with a minimum gap occurring near the Fermi momentum, $k_{F}$. In contrast, in the BEC limit ($\Delta/\epsilon_{F}\gg1$), the quasiparticle dispersion is predominantly governed by the pairing gap, leading to a considerably flattened band dispersion.
Fig \ref{fig:spectral} show these spectral functions, depicting two distinct limits: BCS ($x=0.5$) and BEC ($x=0.9$) based on our slave boson mean field theory of the  ESD t-J model. 
We also illustrate the cases of normal spectral function as well to visualize reversed hierarchy between the Fermi energy and zero-energy in the two opposite limits. 
It is crucial to note that other observables, such as the substantial $\Delta/T_c$ ratio \cite{BEC-BCS2} and density of states asymmetry \cite{BEC-BCS3} also serve as valuable tools for detecting the BEC-BCS crossover. 

The spectral function is 
\begin{eqnarray}
    A(\vec k , \omega )
    &=& -\frac{1}{\pi}\mathrm{Im}
    \left( G_f (\vec k , \omega+ i \eta )\rfloor_{1,1}
    \right)
\end{eqnarray}
with an infinitesimal parameter, $\eta>0$. The fermionic Green's function is defined as, 
\begin{eqnarray}
    G_f(\vec{k}, i\omega_n
    )
    &=& 
    (i\omega_n-\mathcal{H}^{f}_{MF})^{-1},
\end{eqnarray}
with
\begin{eqnarray}
    \mathcal{H}^f_{MF} = 2(\cos k_{x}+\cos k_y)\left( 
    \begin{array}{cc}
        C_{f} & D_{f} \\
        D_{f} & -C_{f} 
    \end{array}
    \right) 
+\delta_{f} I_{2\times 2}, 
\end{eqnarray}
and
\begin{eqnarray}
     C_{f} &=&t\left[ \frac{3}{4} |\langle d \rangle|^2 -\frac{1}{2} |\langle b \rangle|^2\right], \quad
    D_{f} = -2t\sqrt{\frac{3}{8}} \langle  d \rangle \langle b\rangle ,\quad 
    \delta_{f} = -\mu_{0}-\mu.
\end{eqnarray}
In Fig \ref{fig:spectral}, we showed the spectral function setting the parameter $J_{\perp}=5$, $V=1$ and $x=0.5,0.9$ with $\eta=0.001$.

\newpage
\section{Details on DMRG calculation}\label{AE}
In this section we provide more results of our finite and infinite DMRG simulation in the $L_z=2, L_y=1$ configuration.
\subsection{Finite DMRG results}
\begin{figure}[ht]
    \centering
        \includegraphics[width=0.5\textwidth]{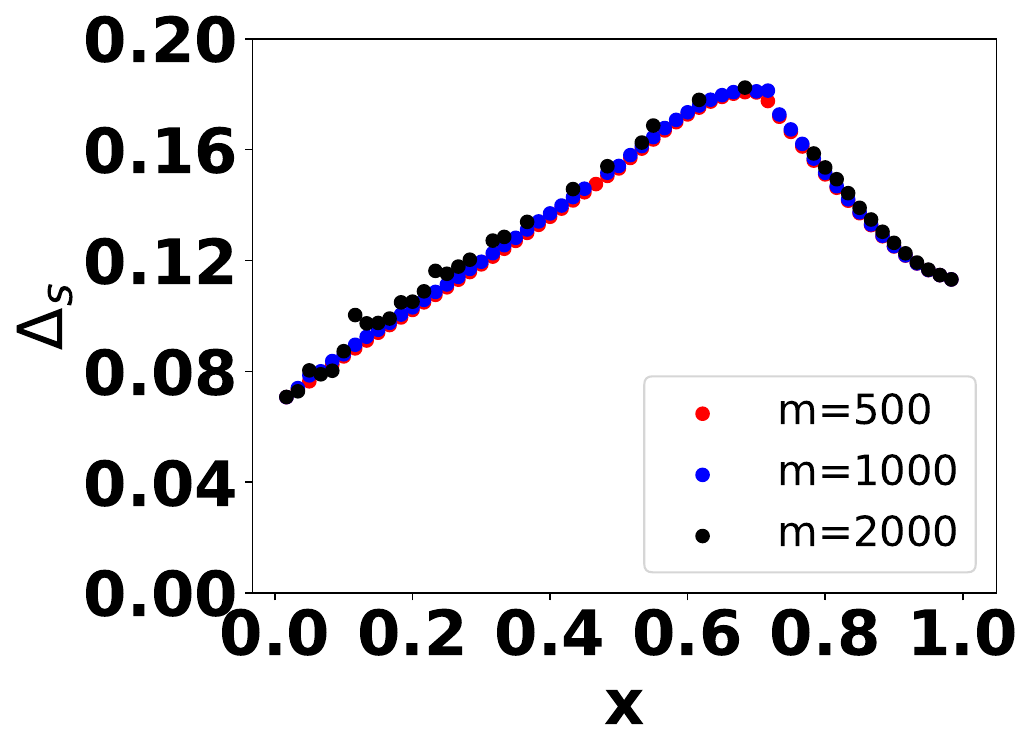}
    \caption{The spin gap for different bond dimensions at $L_x=60$, and the parameters are, $t_\parallel=1$, $t_\perp=0$, $J^{ss}_\parallel=0.5$, $J^{ss}_\perp=1$, $V=0$, and the relation $4J^{dd}=2J^{sd}=J^{ss}$ holds for both $J_\parallel$ and $J_\perp$. We find the DMRG result already converges for $m=2000$.}
    \label{fig:bond_dimension}
\end{figure}

In Fig.~\ref{fig:bond_dimension}, we plot the spin gap for different bond dimensions in type-II t-J model for $L_x=60$, $t_\parallel=1$, $t_\perp=0$, $J^{ss}_\parallel=0.5$, $J^{ss}_\perp=1$, $V=0$, and the relation $4J^{dd}=2J^{sd}=J^{ss}$ for both $J_\parallel$ and $J_\perp$, as in the main text. From Fig.~\ref{fig:bond_dimension}, we find the behavior of the spin gap nearly does not change as we increase the bond dimension, which means the DMRG result already converges for bond dimension $m=2000$.

\begin{figure}[H]
    \centering
    \includegraphics[width=0.9\textwidth]{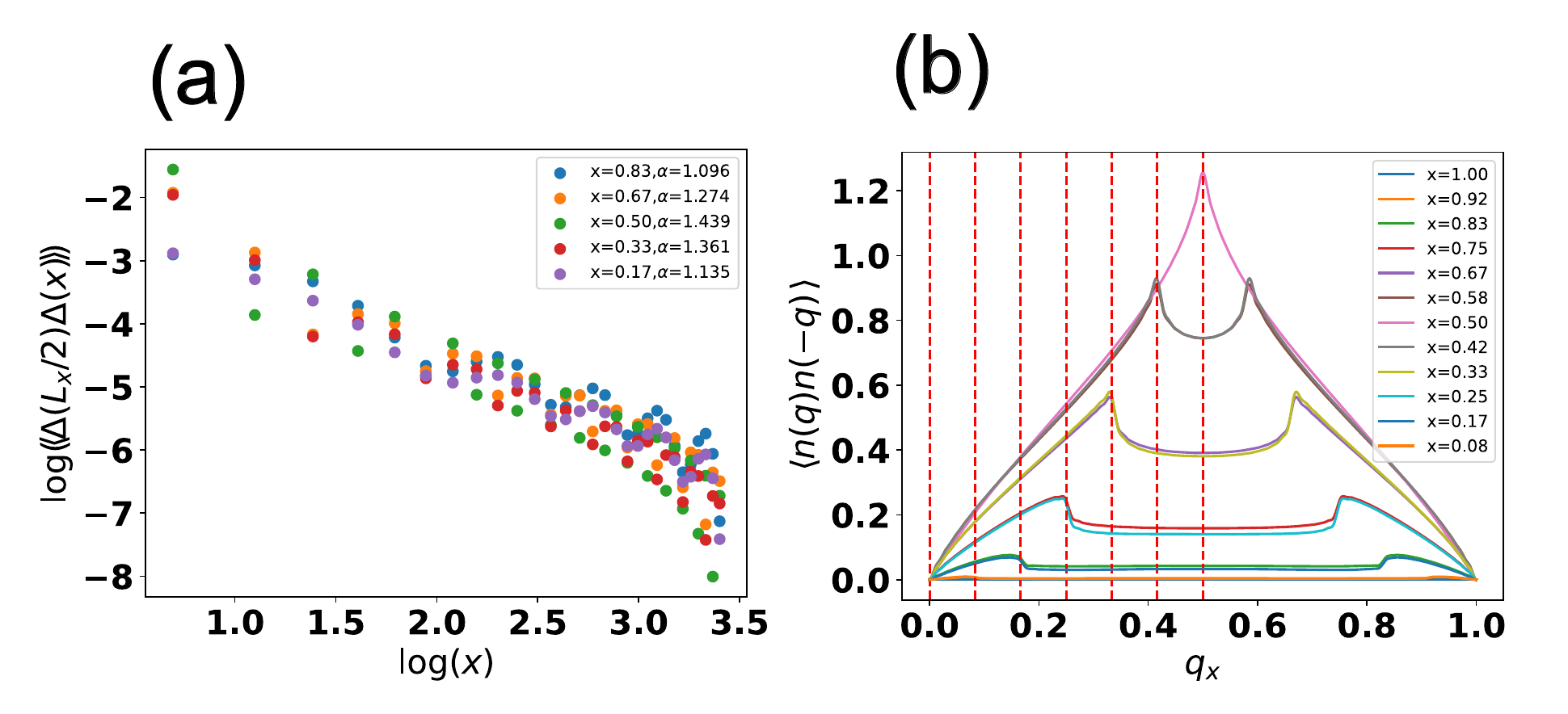}
    \caption{(a) Pairing correlation function $\langle\Delta^\dag(L_x/2)\Delta(x)\rangle$ and (b) density-density correlation function for finite DMRG with $L_x=60$, $t_\parallel=1$, $t_\perp=0$, $J^{ss}_\parallel=0.5$, $J^{ss}_\perp=1$, $V=0$, and the relation $4J^{dd}=2J^{sd}=J^{ss}$ holds for both $J_\parallel$ and $J_\perp$.  In (a) $\Delta=\epsilon_{\sigma\sigma^\prime}c_{t\sigma}c_{b\sigma^\prime}$, and it is plotted in log-scale. Here we can find the power law behavior of the pairing and we fit the exponent by the function $\langle\Delta^\dag(L_x/2)\Delta(x)\rangle=\frac{A}{(L_x/2-x)^\alpha}$. Here we use the data of $x<L_x/2$ to fit the exponent. (b) We can see the peak at different doping is consistent with $q=x=2k_F$ for $x<0.5$ and $q=1-x=2k_F$ for $x>0.5$, in units of $2\pi$, as expected in the Luther-Emery liquid\cite{PhysRevLett.33.589}.}
    \label{fig:pp_dd}
\end{figure}

In Fig.~\ref{fig:pp_dd}, we plot the pairing correlation function $\langle\Delta^\dag(L_x/2)\Delta(x)\rangle$ and the density-density correlation function $\langle n({\bf q})n(-{\bf q})\rangle$ in the type-II t-J model for $L_x=60$, $t_\parallel=1$, $t_\perp=0$, $J^{ss}_\parallel=0.5$, $J^{ss}_\perp=1$, $V=0$, and the relation $4J^{dd}=2J^{sd}=J^{ss}$ holds for both $J_\parallel$ and $J_\perp$. In In Fig.~\ref{fig:pp_dd}(a), the paring is defined as $\Delta(x)=\epsilon_{\sigma\sigma^\prime}c_{t\sigma}c_{b\sigma^\prime}$. In our numerical calculation, we measure $\Delta^\dag(L_x/2)\Delta(x)$. We find the power law scaling of the pairing correlation function, and we further fit the power law exponent from the function $\langle\Delta^\dag(L_x/2)\Delta(x)\rangle=\frac{A}{(L_x/2-x)^\alpha}$. The power law exponent is as expected in the Luther-Emery liquid\cite{PhysRevLett.33.589}, in which the spin is gapped, while the charge remains gapless. In Fig.~\ref{fig:pp_dd}(b), the density operator at each site is defined as $n(x)=n_{2-leg}(2x)+n_{2-leg}(2x+1)$, where $x$ is the site defined in $1$-d and $n_{2-leg}$ is defined in the two-leg ladder. We can see the significant peak in the density-density correlation function, which is related to the Fermi surface. We further find that the peak at $q=x$ for $x<0.5$ and $q=1-x$ for $x>0.5$, in units of $2\pi$, satisfying the Luttinger theorem, which is consistent with the Luther-Emery liquid phase.

\begin{figure}[htbp]
    \centering
    \includegraphics[width=0.9\textwidth]{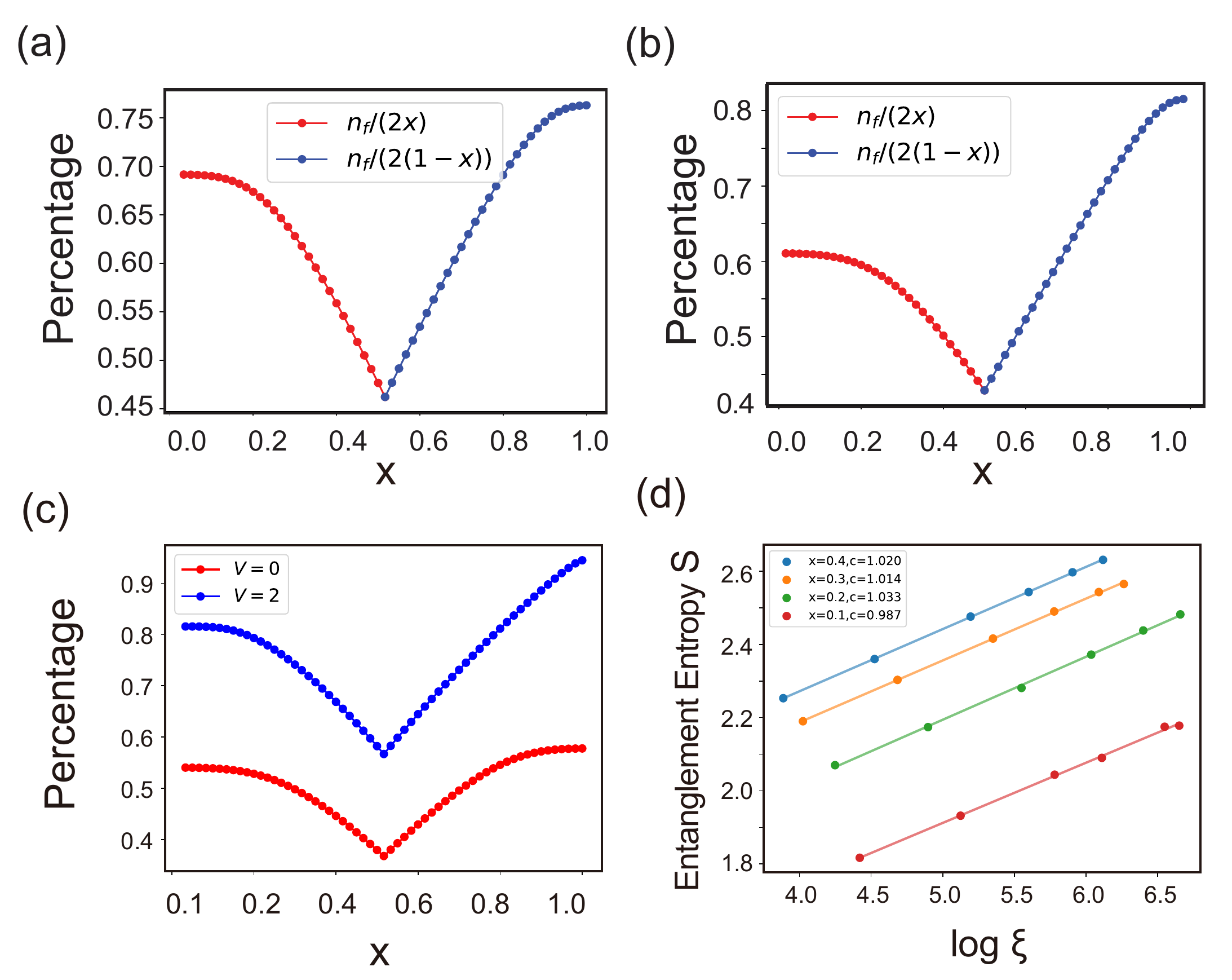}
    \caption{This figure corresponds to Fig.~\ref{fig:dmrg_result} in the main text. (a-b) The percentage of the single-hole, $\frac{n_f}{2x}$ ($x<0.5$), or single-electron, $\frac{n_f}{2(1-x)}$ ($x>0.5$), of the type II t-J model and one-orbital t-J model, respectively. In (a)(b), we use $J_\parallel=0.5$, $J_\perp=1$, and in type-II t-J, we have $J_{\parallel}^{ss}=2J_\parallel^{sd}=4J_\parallel^{dd}=J_\parallel$ and $J_{\perp}^{ss}=2J_\perp^{sd}=4J_\perp^{dd}=J_\perp$. (c) The percentage of $n_f$ substantially increases after introducing the $V=2$ compared to that at $V=0$. Here we use a large $J_\perp=5$. (d) Scaling of the entanglement entropy versus log correlation length yields a central charge $c$, using $S=\frac{c}{6}\log \xi$.  For the iDMRG calculation, we set $J_{\perp}^{ss}=2$ and use four doping concentrations, $x=0.1,0.2,0.3,0.4$.
    For all doping ratio, we find $c=1$ within truncation error of order $10^{-8}$, indicating the evidence of Luther-Emery liquid phase.  
     The presented data is achieved by the bond dimension $m =5\times 10^3$.}
    \label{fig:append_finite_dmrg}
\end{figure}

In Fig.\ref{fig:append_finite_dmrg}, we show the percentage of singlons in type-II t-J model Fig.\ref{fig:append_finite_dmrg}(a) and one-orbit t-J model Fig.\ref{fig:append_finite_dmrg}(b), we find in type-II t-J model the singlons dominate near $x=0$ and $x=1$, corresponding to BCS-like limit, while near $x=0.5$ it is in BEC-like limit with a small percentage of $n_f$. However in Fig.\ref{fig:append_finite_dmrg}(b), in one-orbit t-J model it is always BEC-like. In Fig.\ref{fig:append_finite_dmrg}(c), we show the percentage for a relatively large $J_\perp$ with $V=0$ and $V=2$, we find for $V=0$, it is always BEC-like, but as we increase $V=2$, we find the signature from BCS-like to BEC-like and then back to BCS-like. In Fig.\ref{fig:append_finite_dmrg}(d), we fit the central charge from the relation $S=\frac{c}{6}\log{\xi}$, where $\xi$ is the correlation length, $S$ is the entanglement entropy and $c$ is the central charge. We find the central charge is $c=1$ within error, corresponding to the gapless charge degeree of freedom, which is expected in the Luther Emery liquid.

\begin{figure}[htbp]
    \centering
        \includegraphics[width=0.95\textwidth]{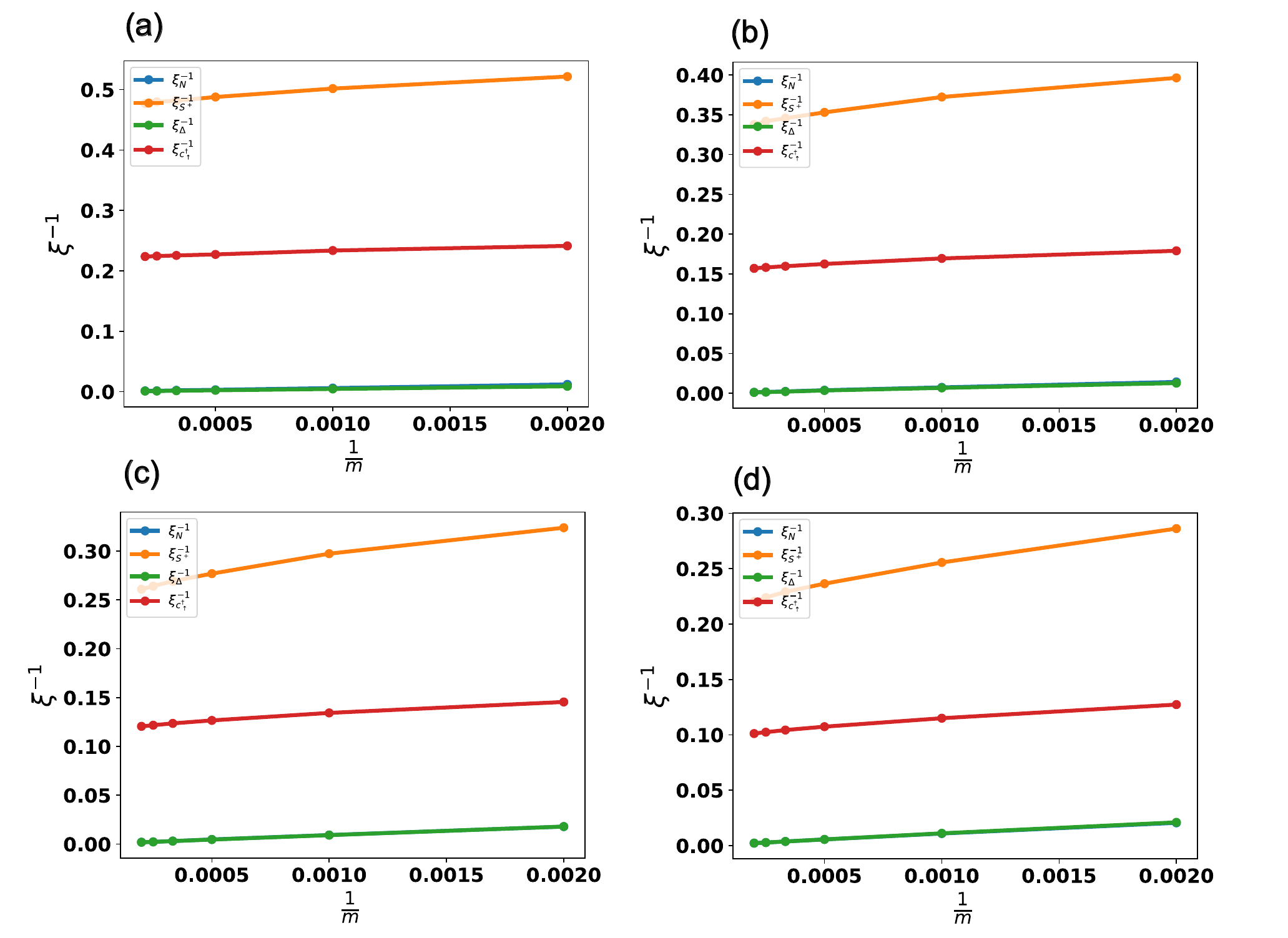}
    \caption{iDMRG results at $t_\parallel=1$, $t_\perp=0$, $J_{\parallel}^{ss}=J_{\parallel}^{dd}=J_{\parallel}^{sd}=0$, $J_{\perp}^{ss}=4J_{\perp}^{dd}=2J_{\perp}^{sd}=2$. (a) (b) (c) (d) correspond to the correlation length at $x=0.1$, $x=0.2$, $x=0.3$, $x=0.4$, respectively. We can see that the correlation lengths of the spin operator and electron operator are finite, while the those of density operator and pairing operator are infinite, corresponding to the gapped spin degree of freedom and gapless charge degree of freedom. Here we get the correlation length from the transfer matrix method\cite{PhysRevB.55.2164,PhysRevLett.75.3537}. The density correlation length $\xi_N$ is from charge section $(\delta Q,\delta S_z)=(0,0)$. The spin correlation length $\xi_{S^\dag}$ is from the charge sector $(\delta Q,\delta S_z)=(0,1)$. The spin correlation length $\xi_{\Delta}$ is from the charge sector $(\delta Q,\delta S_z)=(2,0)$. The electron correlation length $\xi_{c^\dag}$ is from the charge sector $(\delta Q,\delta S_z)=(1,\frac{1}{2})$.}
    \label{fig:idmrg_correlation_length}
\end{figure}

\subsection{Infinite DMRG results}

\begin{figure}[htbp]
    \centering
    \includegraphics[width=0.95\textwidth]{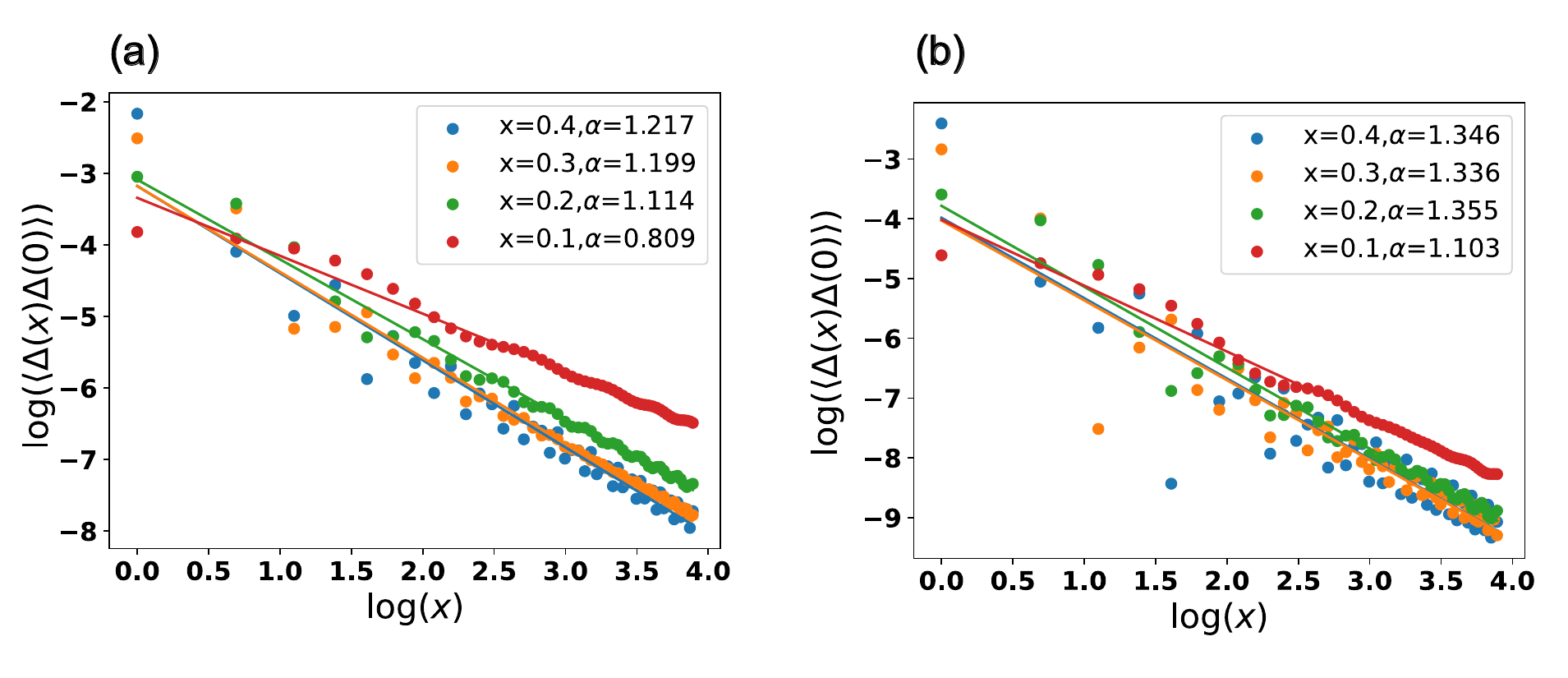}
    \caption{iDMRG results at $t_\parallel=1$, $t_\perp=0$, $J_{\parallel}^{ss}=J_{\parallel}^{dd}=J_{\parallel}^{sd}=0$, $J_{\perp}^{ss}=4J_{\perp}^{dd}=2J_{\perp}^{sd}=2$. (a) and (b) correspond to the pairing correlation function for $V=1$ and $V=2$ in log-scale. We can see the power law scaling, consistent with the Luther-Emery liquid. The exponents are fitted from the relation $\langle\Delta^\dag(x)\Delta(0)\rangle=\frac{A}{x^\alpha}$.}
    \label{fig:idmrg_pairing}
\end{figure}

\begin{figure}[htbp]
    \centering
        \includegraphics[width=0.95\textwidth]{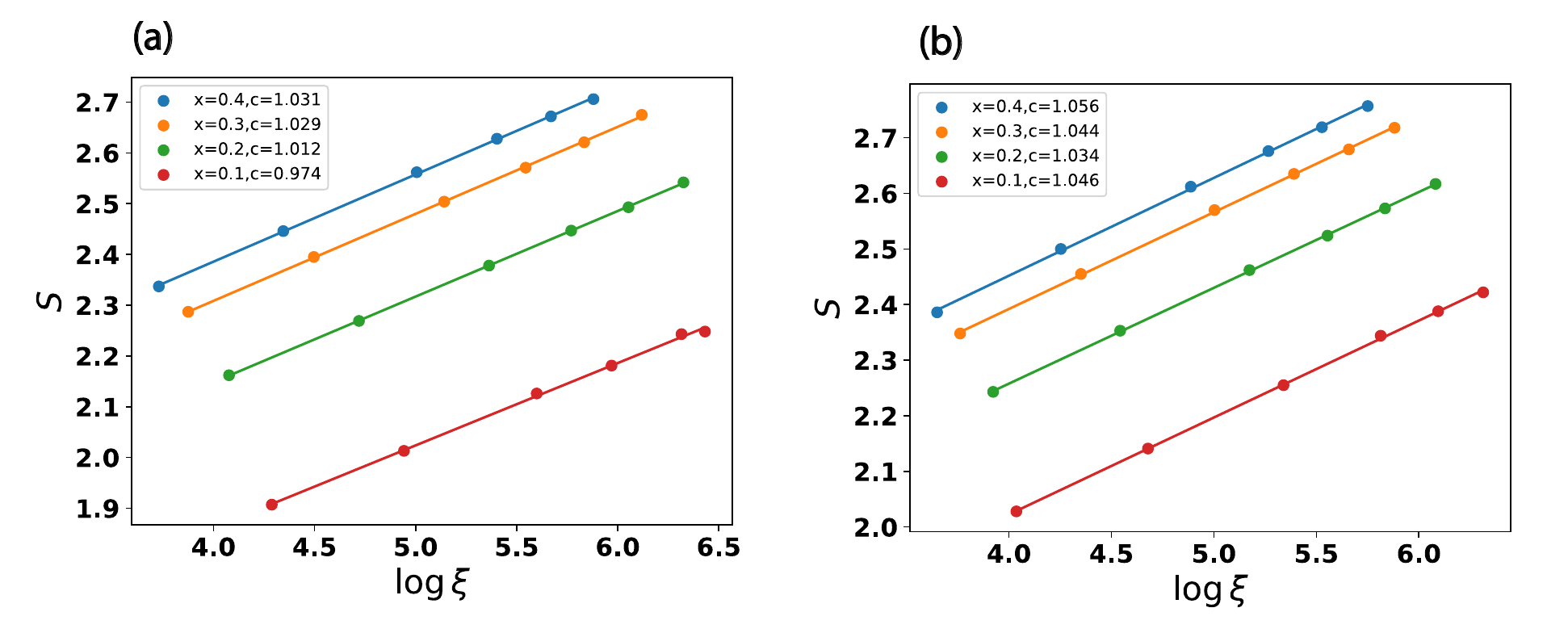}
    \caption{iDMRG results at $t_\parallel=1$, $t_\perp=0$, $J_{\parallel}^{ss}=J_{\parallel}^{dd}=J_{\parallel}^{sd}=0$, $J_{\perp}^{ss}=4J_{\perp}^{dd}=2J_{\perp}^{sd}=2$. (a) and (b) correspond to the central charge for $V=1$ and $V=2$, respectively. We can see the central charge is close to $c=1$, consistent with the Luther-Emery liquid. The exponents are fitted from the relation $S=\frac{c}{6}\log\xi$, where $S$ is the entanglement entropy, $\xi$ is the correlation length, and $c$ is the central charge. The central charge for $V=0$ is in Fig.~\ref{fig:append_finite_dmrg}(d).}
    \label{fig:idmrg_cc_Jp}
\end{figure}

In Fig.~\ref{fig:idmrg_correlation_length} to Fig.~\ref{fig:idmrg_cc_Jp}, we show the infinite DMRG result for a fixed $J_{\perp}^{ss}=4J_{\perp}^{dd}=2J_{\perp}^{sd}=2$ and we increase $V$ from $0$ to $2$. The other parameters are set as $t_\parallel=1$, $t_\perp=0$, $J_{\parallel}^{ss}=J_{\parallel}^{dd}=J_{\parallel}^{sd}=0$. In Fig.~\ref{fig:idmrg_correlation_length}(a)-(d), we get the correlation length of different operators for $V=0$. We find the correlation length is finite for the spin operator and the electron operator, while the correlation length is infinite for the density operator and the pairing operator, in agreement with the gapped spin degree of freedom and the gapless charge degree of freedom. 

In Fig.~\ref{fig:idmrg_pairing}, we plot the pairing-pairing correlation function for $V=1$ and $V=2$. In both cases we still have a power-law decay of the pairing correlation function. The exponent $\alpha$ increases with $V$ because the Luttinger parameter decreases with the repulsive interaction $V$ as expected. But even for $V=2$ the exponent $\alpha$ is still smaller than $2$, indicating a divergent pairing susceptibility.
We have confirmed that the spin correlation length is still finite similar to the results in Fig.~\ref{fig:idmrg_correlation_length}, in agreement with the Luther-Emery liquid.

In Fig.~\ref{fig:idmrg_cc_Jp}(a) and (b), we fit the central charge for $V=1$ and $V=2$ from the relation $S=\frac{c}{6}\log\xi$, where $S$ is the entanglement entropy, $\xi$ is the correlation length, and $c$ is the central charge, and we find the central charge is close to $c=1$, for $V=1$ and $V=2$, consistent with the one gapless mode in the Luther-Emery liquid.

\begin{figure}[htbp]
    \centering
        \includegraphics[width=0.5\textwidth]{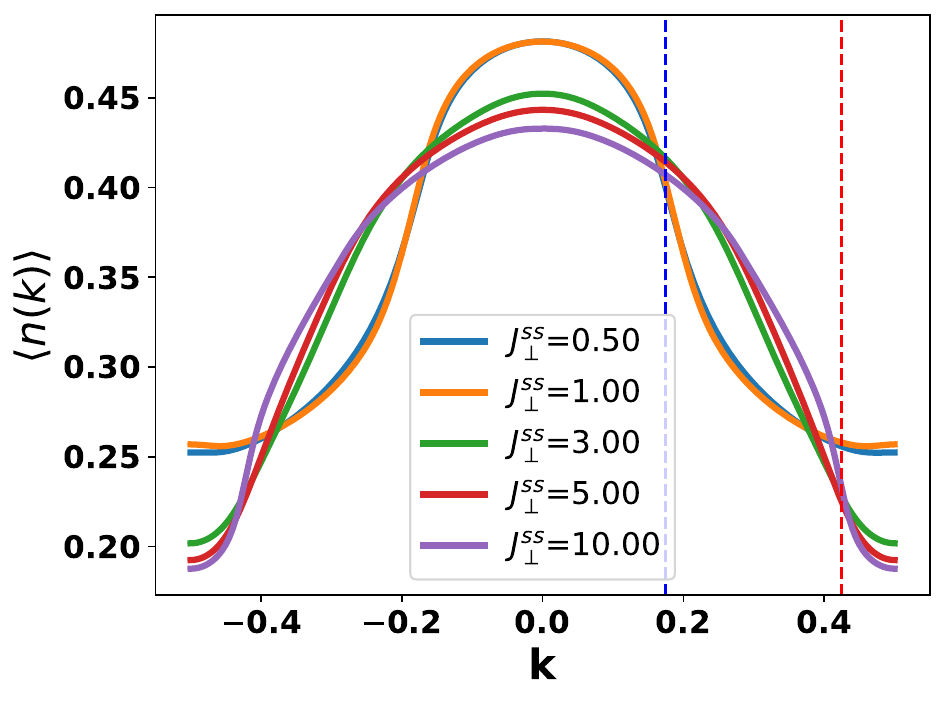}
    \caption{iDMRG results of $n(k)=\langle c^\dag(k)c(k)\rangle$ for fixed $x=0.3$ at $t_\parallel=1$, $t_\perp=0$, $J_{\parallel}^{ss}=0$,$2J_{\parallel}^{dd}=J_{\parallel}^{sd}=0.5$, and $V=100$. The momentum is in units of $2\pi$. There a small to large fermi surface transition as we increase $J_\perp$. The blue and red dashed lines label $k=\frac{1-x}{4}$ and $k=\frac{1}{2}-\frac{x}{4}$ respectively. The large $J_\perp$ side is closer to a small hole pocket centered at $k=\pi$.}
    \label{fig:Nk}
\end{figure}

In Fig.~\ref{fig:Nk}, we plot $n(k)=\langle c^\dag(k)c(k)\rangle$ for fixed $x=0.3$ with a large repulsion $V=100$ to suppress the pairing. In the main text, we show that there is a small to large Fermi surface evolution when increasing $J_\perp$ manifested in the $2k_F$ peaks in the spin-spin correlation function.  Here we show the small to large Fermi surface evolution in the momentum distribution function $n(k)$. The large $J_\perp$ regime is more consistent with a small hole pocket centered at $k=\pi$. This is similar to the sFL phase found in the slave boson theory of the ESD t-J model for the 2D square lattice, where a small hole pocket is centered at $\mathbf k=(\pi,\pi)$.  Note that here the Fermi surface is not very sharp, likely because that the spin gap is not completely suppressed to zero.

\end{document}